\documentclass[
aps,
nofootinbib,
prd,
superscriptaddress,
tightenlines,
notitlepage,
twocolumn,
showpacs
]{revtex4-1}

\usepackage{newtxtext,newtxmath}
\usepackage{enumerate}
\usepackage{bm}
\usepackage{xcolor}
\usepackage{graphicx}	
\usepackage{amsmath}	
\usepackage{float}
\usepackage{multirow}
\usepackage{soul}
\setstcolor{red}
\usepackage{color}
\usepackage{tabularx}
\newcommand{\KIAA}{\affiliation{Kavli Institute for Astronomy and
Astrophysics, Peking University, Beijing 100871, China}}
\newcommand{\DOA}{\affiliation{Department of Astronomy, School of Physics,
Peking University, Beijing 100871, China}}
\newcommand{\LdIT}{\affiliation{Laboratoire des 2 Infinis - Toulouse (L2IT-IN2P3), Universit\'e de Toulouse, CNRS, UPS, F-31062 Toulouse Cedex 9, France}}

\begin{document}

\title{Probing modified gravitational-wave propagation with extreme mass-ratio inspirals}

\author{Chang Liu}\DOA\KIAA\LdIT 
\author{Danny Laghi}\LdIT 
\author{Nicola Tamanini}\LdIT 

\begin{abstract}
Extreme mass-ratio inspirals (EMRIs), namely binary systems composed of a massive black hole and a compact stellar-mass object, are anticipated to be among the gravitational wave (GW) sources detected by the Laser Interferometer Space Antenna (LISA). 
Similarly to compact binary mergers detected by current GW detectors, EMRIs can be used as cosmic rulers to probe the expansion of the Universe.
Motivated by tensions in current cosmological observations as well as by alternative models of dark energy, modified gravity theories can affect the propagation of GWs across cosmological distances, with modifications commonly parametrised in terms of two phenomenological parameters, $\Xi_0$ and $n$.
In this work we adopt a Bayesian approach to constrain for the first time parametrised deviations from General Relativity
using  the loudest simulated EMRIs detected by LISA as dark sirens with a simulated galaxy catalog.
Assuming all the cosmological parameters except $\Xi_0$ are already tightly constrained, our forecasts show that $\Xi_0$ can be constrained to a few percent level (90\% C.I.) with 4 years of LISA observations, unless EMRI detection rates turn out to be closer to current pessimistic expectations.
These results quickly degrade if additional cosmological parameters are inferred simultaneously, but become more robust with an extended LISA observation period of 10 years.
Overall, we find that EMRIs with LISA are better at constraining modified GW propagation than current second-generation ground-based GW detectors, but they will only be comparable to third-generation detectors in the most optimistic scenarios.
\end{abstract}

\maketitle

\section{Introduction}
\label{sec:intro}
Probing the expansion history of our Universe is crucial for testing both general relativity (GR) at large scales and alternatives to $\Lambda$CDM, our current standard cosmological model describing the evolution of the Universe. 
Although $\Lambda$CDM describes very well all the cosmological observations collected so far, growing statistical tensions between measurements taken with different astronomical datasets may constitute a signal for new physics, possibly beyond GR (see e.g.~\cite{Abdalla:2022yfr} for a recent review).
Furthermore, the elusive nature of dark energy, the cosmological entity evoked to explain the observed cosmic acceleration, calls for alternative proposals beyond the simple cosmological constant $\Lambda$ prescribed by $\Lambda$CDM~\cite{Huterer:2017buf}.

Gravitational waves (GWs) from compact binary coalescences provide a unique way to measure the luminosity distance to the source with high precision.
If complementary redshift information can be retrieved, compact binaries can be accurate tracers of the cosmic expansion~\cite{Schutz:1986gp,Holz:2005df}.
This has been clearly demonstrated in several recent works, using different approaches to retrieve redshift information \cite{LIGOScientific:2017adf,soares2019first,gray2020cosmological,LIGOScientific:2021aug,Mastrogiovanni:2021wsd,Gair:2022zsa,PhysRevD.108.042002,gray2023joint}.

Extreme mass-ratio inspirals (EMRIs) are systems consisting of a 
stellar-mass compact object orbiting around a massive black hole with mass in the range $M\in[10^4,10^7] M_{\odot}$, so that they are characterised by a small mass-ratio $q=M_2/M_1 \in [10^{-6},10^{-3}]$~\cite{amaro2007intermediate}. These systems are one of the target GW sources that will be detected by the Laser Interferometer Space Antenna (LISA)~\cite{Babak:2017tow,LISA:2017pwj} and, if used as dark sirens, i.e.~by employing complementary redshift information collected by cross-correlating sky-localisation with galaxy catalogs, they have the potential to place stringent constraints on cosmological parameters within the framework of GR~\cite{macleod2008precision,Laghi:2021pqk} and complement other LISA cosmological measurements~\cite{Tamanini:2016zlh,Corman:2021avn,Toscani:2023gdf,Speri:2020hwc,LISACosmologyWorkingGroup:2019mwx,Cai:2017yww,Caprini:2016qxs,LISACosmologyWorkingGroup:2022jok}.

GWs can be used to study general extensions of gravity that differ from GR on cosmological scales, a possibility that could help explain (some of) the cosmological tensions as well as the origin of dark energy~\cite{saltas2014anisotropic,lombriser2016breaking,nishizawa2018generalized,Belgacem:2017ihm,arai2018generalized,wolf2020standard}.
Modified gravity theories can produce deviations at the level of cosmological perturbations, giving rise to modifications in the tensor perturbations which can affect the luminosity distance measured by standard sirens at cosmological distances~\cite{Nishizawa:2017nef,Arai:2017hxj,Belgacem:2017ihm,Belgacem:2018lbp}.
One possible general way of describing these deviations from GR, which mainly originate from an additional damping of the GW amplitude, is by introducing two phenomenological parameters ($\Xi_0$, $n$) in the luminosity distance-redshift relation~\cite{Belgacem:2017ihm,Belgacem:2018lbp, Lagos:2019kds, Zhu:2023wci}.

Recent measurements on $\Xi_0$ from the GW data of GWTC-3 binary black holes~\cite{LIGOScientific:2021djp}, the latest catalog of GW detections by the LIGO-Virgo-KAGRA (LVK) Collaboration, yield order one constraints at the 90\% confidence intervals (CIs) level \cite{Finke:2021aom,Mancarella:2022cgn}, over an expected GR value of $\Xi_0 = 1$.
Bright siren constraints on $\Xi_0$ from GW170817, the only multi-messenger binary neutron star merger detected by the LVK Collaboration so far, are at least one order of magnitude worse~\cite{Belgacem:2018lbp,Mastrogiovanni:2020mvm}.

In this work, we explore LISA capability to constrain possible deviations from GR at cosmological distances for 3 different EMRI populations, 2 EMRI waveforms, and 2 different mission durations.
In particular, we study the constraints on modified GW propagation effect and provide limits on cosmological parameters $\Omega=\{h,\ \Omega_M,\ \Xi_0,\ n\}$.
The constraints on $\Xi_0$ range from 1.3\% in the most optimistic scenario (M6, 10yrs, AKK) to no constraint in the most pessimistic scenario (M5, 4yrs, AKS).
We also report updated EMRI forecasts for the $\Lambda$CDM model.

The paper is organised as follows.
In Sec.~\ref{sec:modGWpro} we introduce the ($\Xi_0$, $n$) parametrisation to describe deviation from GR in the cosmological propagation of GWs.
In Sec.~\ref{sec:method} we outline our simulated EMRI observations and Bayesian methodology employed to constrain cosmological parameters.
In Sec.~\ref{sec:results} we present our results under different assumptions for the EMRI population, the waveform model, and the LISA observation time period.
Finally, we compare our results with the current literature and draw our conclusions in Sec.~\ref{sec:conclusion}.


\section{Modified GW propagation}
\label{sec:modGWpro}

In extensions of the $\Lambda$CDM model which generate a dynamical dark energy (DE) sector, the  luminosity distance to standard sirens is:
\begin{equation}
d_L=\frac{c}{H_0}(1+z) \int_0^z \frac{d \tilde{z}}{\sqrt{\Omega_M(1+\tilde{z})^3+\rho_{\mathrm{DE}}(\tilde{z}) / \rho_0}}\,,
\label{eq:dLEM}
\end{equation}
where $H_0$ is the value of the Hubble constant today, $\Omega_M$ is the present fraction of matter energy density, $\rho_{\rm{DE}}$ is the DE energy density, $\rho_0$ is the critical energy density defined as $\rho_0=3 H_0^2 /(8 \pi G)$, and $c$ is the speed of light.
In $\Lambda$CDM, dark energy is nothing but the cosmological constant $\Lambda$, meaning $\rho_{\rm DE} / \rho_0 = \Omega_\Lambda = 1 - \Omega_M$ (assuming flatness), where the constant $\Omega_\Lambda$ is the present fraction of vacuum energy density.

In modified gravity theories, both the cosmological background evolution
and the dynamics of cosmological perturbations can deviate from what the standard $\Lambda$CDM model predicts.
The modification to the background evolution is usually recast in a (dynamical) change of the dark energy equation of state, providing a non-constant $\rho_{\rm DE}(z)$. 
At the perturbation level instead, both scalar perturbations, which affect the propagation of photons and the evolution of large-scale structures, and tensor perturbations, which correspond to GWs propagating through cosmological distances, are modified by beyond-$\Lambda$CDM corrections. 
Under very general conditions, in most modified gravity scenarios, GW propagation through a homogeneous and isotropic universe obeys the following equation of motion \cite{Nishizawa:2017nef,Arai:2017hxj}:
\begin{equation}
\tilde{h}_A^{\prime \prime}+2 \mathcal{H}[1-\delta(\eta)] \tilde{h}_A^{\prime}+c_{\rm GW}^2k^2 \tilde{h}_A=0\,,
\end{equation}
where $\tilde{h}_A(\eta, k)$ is the Fourier transformed GW amplitude and ${A}$ denotes the polarization mode $+$ or $\times$, $\mathcal{H}=a'/a$ is the conformal Hubble parameter, while the prime $^\prime$ denotes differentiation with respect to the conformal time $\eta$.
Here the function $\delta(\eta)$ modifies the friction term, changing how the GW amplitude dampens across cosmological distances with respect to GR (where $\delta(\eta)=0$).
In the third term, the coefficient $c_{\rm GW}$ is the speed of propagation of GWs~\cite{deRham:2018red}, which in GR equals the speed of light: $c_{\rm GW} = c$.
This parameter has already been constrained at the level $\frac{1-c_{\rm GW}}{c}\le 3 \times 10^{-15}$ in the LIGO-Virgo frequency band ($\sim$kHz) by the joint observation of GW170817 and GRB170817A~\cite{LIGOScientific:2017zic} and could be tested also in the LISA frequency band~\cite{baker2022measuring}.
In what follows we will therefore assume $c_{\rm GW} = c$.

In this study we will assume that DE is represented by the cosmological constant and focus on the friction term: when $\delta(\eta)\neq 0$, the luminosity distance extracted from GW observations $d_{L}^{\rm GW}$ deviates from the standard luminosity distance $d_L^{\rm EM}$ measured by electromagnetic (EM) observations, which is given by Eq.~\eqref{eq:dLEM} with $\rho_{\rm DE} / \rho_0 = \Omega_\Lambda$.
The relation between $d_{L}^{\rm GW}$ and $d_{L}^{\rm EM}$ can be expressed as:
\begin{equation}\label{eq:delta_z}
d_L^{\mathrm{GW}}(z)=d_L^{\mathrm{EM}}(z) \exp \left\{-\int_0^z \frac{d z^{\prime}}{1+z^{\prime}} \delta\left(z^{\prime}\right)\right\} \,.
\end{equation}
Therefore we see that GWs can test modified theories of gravity through the function $\delta(z)$.
For most modified gravity theories, it is usually difficult to reconstruct a full relation between redshift and $\delta(z)$ from the data.
Nevertheless a convenient phenomenological two-parameter parametrization has been proposed in the literature~\cite{Belgacem:2017ihm}:
\begin{equation}
    \frac{d_L^{\mathrm{GW}}(z)}{d_L^{\mathrm{EM}}(z)}=\Xi_0+\frac{1-\Xi_0}{(1+z)^n}\,,
    \label{eq:Xi0}
\end{equation}
where $\Xi_0$ and $n$ are the only two beyond-$\Lambda$CDM parameters of the model.
The ratio ${d_L^{\mathrm{GW}}(z)}/{d_L^{\mathrm{EM}}(z)} \rightarrow 1$ as $z \rightarrow 0$, as we expect no deviations in the local Universe, while ${d_L^{\mathrm{GW}}(z)}/{d_L^{\mathrm{EM}}(z)} \rightarrow \Xi_0$ as $z \rightarrow \infty$, since the main hypothesis behind most theories assumes that GW deviations should predominantly affect the recent cosmological epoch, thus at large $z$, $\delta(z) \rightarrow 0$, and from Eq.~\eqref{eq:delta_z} the ratio ${d_L^{\mathrm{GW}}(z)}/{d_L^{\mathrm{EM}}(z)}$ should deviate from unity and tend to a constant value as redshift increases.
$\Xi_0$ is a more important parameter to be constrained, as it determines by how much the ratio $d_L^{\mathrm{GW}}(z)/{d_L^{\mathrm{EM}}(z)}$ deviates from 1 at high redshift.
The parameter $n$ instead controls how fast this ratio changes from 1 at $z=0$ to $\Xi_0$ at $z=\infty$.
Assuming a transition happens rapidly at late-time, the rate $n$ at which this transition takes place comes as a secondary issue with respect to the depth $\Xi_0$ of the transition itself.
However, it is worth noting that GR can be recovered also in the limit $n \rightarrow 0$, so we may expect some potential degeneracy between the parameters $\Xi_0$ and $n$ when the true cosmology of our simulations is based on GR.

Our fiducial values for the $\Lambda$CDM  parameters are $h = H_0 / (100 \,{\rm km/s/Mpc}) =0.673$, $\Omega_M=0.315$,
which are based on the true cosmology used to produce the simulated galaxy catalog that we use in our study \cite{villalba19}.
For the modified gravity parameters, we will explore the GR case and two deviations from it, namely $\Xi_0=0.9, 1.0, 1.2$, while in all our non-GR data sets we will assume $n=2$, which is close to the prediction made in some non-local gravity models, see e.g.~\cite{Finke:2021aom}.
This specific value of $n$ is substantially irrelevant for the purpose of testing our statistical methodology with EMRIs; moreover, we will also consider scenarios where $n$ is let vary during the parameter inference.


\section{Method}
\label{sec:method}

We use the dark sirens method in a Bayesian statistical framework to estimate how EMRI populations detected by LISA could constrain the cosmological parameters $\Omega=\{h,\ \Omega_m,\ \Xi_0,\ n\}$.
In the most general case, we assume uniform priors for the following cosmological parameters: $h\in[0.6, 0.76]$, $\Omega_M\in[0.04, 0.5]$, $\Xi_0\in[0.6,2.0]$, $n\in[0.0, 3.0]$.
The lower prior on $\Xi_0$ is chosen for the sake of computational cost.
In fact, lower values of $\Xi_0$ produce strong correlations in the distance-redshift relation, Eq.~\eqref{eq:Xi0}, which in turn make the dark siren statistical inference more complex.
Our approach closely follows the one adopted in \cite{Laghi:2021pqk}.
In what follows we describe our simulated data set step by step.

\subsection*{EMRI population catalogs}
EMRI rates depend on several astrophysical assumptions about the MBH mass function and its redshift evolution, the fraction of MBHs hosted in dense stellar cusps, and binary parameters such as the eccentricity and the mass of the compact object (for more details, see~\cite{Babak:2017tow}). The EMRI study in~\cite{Babak:2017tow} considered different assumptions which led to a set of EMRI astrophysical models (cf. their Table I), spanning several orders of magnitude in EMRI rates. The sources of each of these models have then been analysed with a Fisher Matrix (FM) approach to obtain detection catalogs with estimates of the LISA parameter estimation accuracy for the detected sources, where a detection signal-to-noise ratio (SNR) threshold of 20 was originally chosen. In this study, in order to account for the uncertainty in rate estimates, we focus on three of the EMRI population models first analysed in~\cite{Babak:2017tow}, which cover different EMRI rates.
In particular, keeping the original notation introduced in~\cite{Babak:2017tow}, we consider a fiducial model (M1), a pessimistic model (M5) and an optimistic model (M6).
The FM analysis in~\cite{Babak:2017tow} made use of two approximate analytical waveform models to estimate the waveform generated by EMRIs, the so-called Analytical Kludge Schwarzschild (AKS) and Analytical Kludge Kerr (AKK) waveforms~\cite{barack2004lisa}. 
These two waveform models are truncated at the Schwarzschild and the Kerr last stable orbits of the MBH, respectively, which are used as proxies for the plunge. 
In fact, the choice of various plunge conditions establishes how close the compact object can approach the MBH horizon, thereby defining the bandwidth and SNR of the signal; thus, AKK waveforms in general produce more orbits than the AKS.
As a rule of thumb, the AKS waveform underestimates the signal's
strength, while the AKS significantly overestimates it as the Analytical Kludge models become increasingly inaccurate near the horizon~\cite{Babak:2017tow}.
In this study we are interested in a subset of EMRI extrinsic parameters, namely the sky localisation, since the smaller the localisation error volume, or error box, of the GW source is, the fewer host galaxies will fall within this region, making the events more informative for cosmology.

The FM catalogs for each model in~\cite{barack2004lisa} provide the measured EMRI luminosity distance $d_L^{\rm GW}$, longitude $\phi_{\rm GW}$, and colatitude $\mu_{\rm GW}$ (where $\mu_{\rm GW}=\cos(\theta_{\rm GW})$), with their uncertainties $\sigma_{d_L^{\rm GW}}$, $\sigma_{\phi_{\rm GW}}$, and $\sigma_{\mu_{\rm GW}}$,
and the sky-location covariance matrix $\Sigma_{2\times2}$. These quantities represent the only GW data $D$ that we need to build our simulated set of dark siren observations.
The original EMRI catalogs of \cite{Babak:2017tow} contain events collected over a 10 year LISA observation period.
We obtain catalogs for 4 years of LISA observation by randomly drawing 4/10 of the total number of 10 year detections from these original catalogs.

\subsection*{Galaxy Catalog}
The next step in building our catalog of dark standard sirens is to choose a simulated galaxy catalog containing information about the redshift distribution of the galaxies.
In this study we use a galaxy lightcone built adopting the same method developed in~\cite{villalba19}, which transforms the outputs of \texttt{L-Galaxies}~\cite{henriques2015galaxy}, a state-of-the-art semi-analytical model applied to the Millennium simulation~\cite{springel2005cosmological}, into a lightcone with galaxies and their physical properties. Our lightcone, which has been used also in~\cite{muttoni2023dark}, covers one octant of the sky and has a resolution of $M_{\star} > 10^{10} M_{\odot}$. As a result of this resolution limit, we may miss lower-mass galaxies that could potentially be host to EMRIs. In practice, a lower mass cut-off would mean that a larger number of galaxies would fall into each EMRI localisation error volume, potentially worsening our cosmological parameter estimates. However, as shown in Appendix A1 of~\cite{Laghi:2021pqk}, this limit is not expected to relevantly affect the cosmological inference that could be obtained with the dark siren approach, thus our conclusions should not be affected by the lightcone resolution.
The information we need for each galaxy is the cosmological redshift $z_{\rm gal}^{\rm cos}$, (that is only used in the error box construction), the observed redshift $z_{\rm gal}^{\rm obs}$ (from which, using Eq.~\eqref{eq:dLEM}, we can compute the luminosity distance measured from EM observations $d_L^{\rm gal, EM}$, and it is thus used in both in the error box construction and in the likelihood), and the right ascension $\phi_{\rm gal}$ and cosine of the colatitude $\mu_{\rm gal}=\cos(\theta_{\rm gal})$.

\subsection*{Errorbox construction}
For the construction of the error boxes, we closely follow the procedure outlined in~\cite{Muttoni:2021veo}, with the difference that here we are dealing with the framework of modified GW propagation theory.
In practice, for each GW event we construct localisation error volumes, or error boxes, in redshift space by first converting the lower and upper 3$\sigma$ uncertainty bounds on $d_L^{\rm GW}$, as provided by the GW measurements, adding in quadrature the weak lensing contribution to the measurement uncertainty:
\begin{equation}\label{eq:sigma_dl}
{\sigma_{d_L}}=\sqrt{\left({\sigma_{d_L^{\rm GW}}}\right)^2+\left({\sigma_{d_L^{\rm WL}}}\right)^2}\,,
\end{equation}
where the weak lensing error $\sigma_{d_L^{\rm WL}}$ is estimated using the model described in~\cite{Tamanini:2016zlh}.

The first step in constructing an EMRI localisation error volume is to select the true host of the GW event. Assuming the true cosmological model, i.e.,  the fiducial cosmological parameters, we compute the redshift range $[z^-_{\rm tr}$, $z^+_{\rm tr}]$ from the $d_L^{\rm GW}-3\sigma_{d_L}$ and $d_L^{\rm GW}+3\sigma_{d_L}$ values, respectively.
It is worth noting that employing 2$\sigma$ errors, instead of 3$\sigma$ errors, would lead to more EMRI events within the lightcone redshift boundary and fewer host candidates in a given error box, which would greatly help our numerical analysis, but would only provide results with a maximum precision of few \% invalidating our most optimist results.
We list all the galaxies having $z^{\rm cos}_{\rm gal} \subset [z^-_{\rm tr}$, $z^+_{\rm tr}]$ and for each of them, we compute the corresponding modified gravity luminosity distance $d_L^{\rm gal, GW}$ using Eq.~\eqref{eq:Xi0}, where $d_L^{\rm gal,EM}$ is computed using Eq.~\eqref{eq:dLEM} and $z_{\rm gal}^{\rm obs}$.
$d_L^{\rm gal, GW}$ represents the luminosity distance measured by the GW observation when the galaxy is the host: even though this quantity is not directly observed (galaxies are observed in the EM spectrum, where the luminosity distance is given by Eq.~\eqref{eq:dLEM}), it is required to associate in a consistent way a galaxy host with the EMRI, for which we only have observation of its modified luminosity distance $d_L^{{\rm GW}}$, Eq.~\eqref{eq:Xi0}. These values are only used to extract a random galaxy with probability given by $\left.\mathcal{N}\left(d_L^{\rm GW}, \sigma_{d_L}^2\right)\right|_{d_{L}^{\rm gal, GW}}$, which will be the true host of the GW source. The sky coordinate of the true host is then labeled as $\{\mu_{\rm tr}, \phi_{\rm tr}\}$.
The next step is to displace the center of the localization error volume, corresponding to the most probable sky coordinates measured from the GW observation.
This is done to simulate the fact that due to the random nature of the observation noise we do not expect the EMRI to be placed exactly at the center of the measured localisation error volume.
We build a 2D Gaussian probability distribution $\mathcal{N}\left([\mu_{\rm tr}, \phi_{\rm tr}], \Sigma_{2\times2}\right)$ centered at the true host coordinates, from which we randomly draw a value that will be the error box center of the GW observation, and denote it as $\{\mu_{\rm GW}, \phi_{\rm GW}\}$. 
Next, we calculate the redshift boundaries of the error boxes. Considering all the values that the cosmological parameters can assume within their prior ranges, as well as the measured luminosity distance range $[d_L^{\rm GW}-3\sigma_{d_L}, d_L^{\rm GW}+3\sigma_{d_L}]$, we use the modified GW relation Eq.~\eqref{eq:Xi0} to compute the redshift boundaries that will contain all the potential host galaxies from the galaxy catalog that could potentially host the EMRI, $[z^-, z^+]$.
We then assume a rms peculiar velocity of the galaxy ${v_p} = 700$ km/s (which is estimated from our simulated light cone) and transform it to a galaxy peculiar velocity uncertainty with $\sigma_{\rm pv}(z) \simeq {v_p}(1+z)/c = 0.0023(1+z)$. 
Finally, we select all the galaxies that satisfy $z^--\sigma_{\rm pv}(z^-) <z_{\rm gal}^{\rm obs}< z^++\sigma_{\rm pv}(z^+)$, $\phi_{\rm GW}-3\sigma_{\phi_{\rm GW}}<\phi_{\rm gal}<\phi_{\rm GW}+3\sigma_{\phi_{\rm GW}}$ and $\mu_{\rm GW}-3\sigma_{\mu_{\rm GW}}<\mu_{\rm gal}<\mu_{\rm GW}+3\sigma_{\mu_{\rm GW}}$. The galaxies of the $i$-th EMRI detection are labeled by $j \ (j=1, ...,N_{\rm gal}^i)$. Each candidate host is given a weight according to its position relative to the center of the error box, $w_j \propto \mathcal{N}\left([\mu_{\rm GW}, \phi_{\rm GW}], \Sigma_{2\times2}\right)|_{(\mu_{\rm gal}^j, \phi_{\rm gal}^j)}$, which will then be used in the likelihood. As a final step, we exclude the events whose error box contains more than 5000 galaxies as these events are expected to be largely uninformative for cosmological inference. This choice saves considerable computational time and we have numerically verified that it has little to no loss in parameter estimation accuracy.

\subsection*{Bayesian inference}
According to Bayes' theorem, the posterior probability distribution of the parameters $\Omega$ can be written as~\cite{DelPozzo:2011vcw, Laghi:2021pqk}:
\begin{equation}
p(\Omega \mid D \mathcal{H} I)\propto p(\Omega \mid \mathcal{H} I) p(D \mid \Omega \mathcal{H} I)\,,
\label{eq:posterior}
\end{equation}
where $D$ is the data from GW observations coming from the FM analysis, $\mathcal{H}$ is the assumed modified gravity cosmological model, and $I$ is the background information relevant for the analysis.
$p(\Omega \mid \mathcal{H} I)$ is the prior on $\Omega$, which we assume as a uniform distribution for each parameter.
$p(D \mid \Omega \mathcal{H} I)$ is the likelihood of $N$ EMRI events which, assuming they could be studied and resolved individually, can be written as the product of $N$ single-event likelihoods:
\begin{equation}
p(D \mid \Omega \mathcal{H} I)=\prod_{i=1}^N p(D_i \mid \Omega \mathcal{H} I)\,.
\end{equation}
After marginalising over $d_L$, the single-event likelihood can be written as:
\begin{equation}
p(D_i \mid \Omega \mathcal{H} I)=\int dz_{\rm GW}
p\left(D_i \mid z_{\rm GW}\Omega \mathcal{H} I\right)
p(z_{\rm GW} \mid \Omega \mathcal{H} I),
\label{eq:lklhd_single}
\end{equation}
where
\begin{equation}
p(z_{GW} \mid \Omega \mathcal{H} I) \propto 
\sum_{j=1}^{N_{\rm gal}^i} \frac{w_j}{\sigma_{\rm pv}({z_{{\rm gal}, j}^{\rm obs}})} 
\exp \left\{-\frac{1}{2}\left[\frac{z_{{\rm gal}, j}^{\rm obs}-z_{\rm GW}}{\sigma_{\rm pv}({z_{{\rm gal}, j}^{\rm obs}})}\right]^2\right\}\,,
\end{equation}
and
\begin{equation}
p(D_i \mid z_{\rm GW}\Omega \mathcal{H} I)\propto  
\exp \left\{-\frac{1}{2}\left[\frac{\hat{d}_L^{\rm GW} - d_L^{\rm GW}(\Omega,z_{\rm GW})}{\sigma_{D_L}}\right]^2\right\}\,,
\end{equation}
where $\hat{d}_L^{\rm GW}$ is the best luminosity distance estimate by LISA (coming from the FM analysis) and $d_L^{\rm GW}(\Omega,z_{\rm GW})$ is obtained from the modified luminosity distance-redshift relation, Eq.~\eqref{eq:Xi0}.
For each event, we integrate the likelihood Eq.~\eqref{eq:lklhd_single} numerically and infer the posterior distribution Eq.~\eqref{eq:posterior} with \texttt{cosmoLISA}~\cite{cosmoLISA}, a public code used for cosmological inference with LISA GW sources. The posterior distribution is explored with \texttt{raynest}~\cite{raynest}, a parallel nested sampling algorith based on the package \texttt{ray}~\cite{ray}, an open-source package that allows for parallel processing of jobs.
Compared to~\cite{Laghi:2021pqk}, in this study we use an updated version of \texttt{cosmoLISA}, where the major optimisation concerns the numerical marginalisation over the GW redshift in Eq.~\eqref{eq:lklhd_single}. In particular, we do not obtain redshift posterior samples
for each EMRI event along with the cosmological parameters
as in [14], but marginalise over the redshift numerically while computing the likelihood, therefore reducing the parameter space to be explored and  the computational time. We note that, from the formal point of view of nested sampling, the two approaches give equivalent results, with the
advantage that in the new approach the parameter space to be sampled is smaller and only defined by the cosmological parameters.
This results in an acceleration of the likelihood and nested sampling, and a version of the code on average 10 times faster than its predecessor when used on a computational cluster and with a similar number of GW events.

The inference is done only on the most informative EMRIs with a signal-to-noise ratio (SNR) threshold that is customarily set to 100. We further impose a redshift cut at $z=1$ for the upper bound of the localisation volume in redshift space.
These conditions ensure that selection effects due to the limited sensitivity of the GW detector, the modelling of the underlying source population, and the finite redshift coverage of the galaxy catalogue do not introduce significant bias in the inference.
In analogy to~\cite{Laghi:2021pqk}, we performed several numerical tests to check that our results don't get biased with our chosen SNR threshold and verified numerically that such biases become relevant only if sources of lower SNR are considered. 
A detailed treatment of selection effects (assuming a complete galaxy catalog, as in our case, so that EM selection effects do not affect the study) would require modelling the EMRI detection probability for LISA, and the assumption of hyper-parameters describing both the EMRI population distribution (like the source-frame masses of the systems) and the merger rate redshift evolution, which are needed to compute the expected number of detectable GWs. 
An appropriate statistical framework for this approach would require a fully hierarchical inference approach~\cite{mandel2019extracting,vitale2022inferring}, allowing joint inference of the cosmological, population and rate hyper-parameters~\cite{Mastrogiovanni:2021wsd,Gair:2022zsa,PhysRevD.108.042002,gray2023joint}, which requires a computational cost that goes beyond our current numerical resources, and whose investigation we thus defer for future works.
Our selections in SNR and redshift do not pose a severe limitation to our forecasts, as low-SNR detections are generally not well-localised and are not expected to be informative dark sirens.


\section{Results}
\label{sec:results}

We report results for the three EMRI population models M1 (fiducial), M5 (pessimistic), M6 (optimistic) for both 4 and 10 years of LISA observation and using both waveform models at our disposal (AKS and AKK).
We constrain the cosmological parameters $h,\ \Omega_m,\ \Xi_0,\ n$ either separately or collectively, for a total of five different cosmological parameter scenarios: i) $h+\Omega_M+\Xi_0+n$, ii) $h+\Omega_M+\Xi_0$, iii) $h+\Xi_0$, iv) $\Xi_0+n$, and v) $\Xi_0$ only.
We will additionally analyse the GR $\Lambda$CDM scenario ($h+\Omega_M$) to compare our findings with the EMRI dark siren study in~\cite{Laghi:2021pqk}.

For each cosmological scenario, EMRI population model, time of observation, and waveform model, we realise 5 independent realizations of the error boxes by following the procedure outlined in Sec.~\ref{sec:method}.
Each realisation is based on the same LISA catalogs of detected EMRIs within a given cosmological scenario, but the resulting localisation error volume for a given EMRI will generally be different in each realisation, since the EMRI extrinsic parameters (in particular the sky localisation angles) are randomised so that their host galaxies differ in each realisation.
We then perform the cosmological inference on each of the 5 realisations and rank them according to their marginalised 90\% C.I.~on $\Xi_0$, which is the parameter present in all our cosmological scenarios, selecting the median realisation to represent the forecasts in each scenario.

We first show results for 4 years of LISA observation: in Sec.~\ref{subsec:Xi=1} and Sec.~\ref{subsec:Xi=0912} we respectively assume a true value of $\Xi_0 =1.0$, corresponding to the GR limit (see Table~\ref{tab:10}), and deviations from GR with $\Xi_0 =0.9$ (see Table~\ref{tab:09}) and $\Xi_0 =1.2$ (see Table~\ref{tab:12}).
Subsequently, in Sec.~\ref{subsec:10yr} we provide results for 10 years of LISA observation with an injected value $\Xi_0 = 1.0$ (see Table~\ref{tab:10yr10}).
Lastly, in Sec.~\ref{subsec:LCDM} we present $\Lambda$CDM results for comparison with the literature~\cite{Laghi:2021pqk} (see Table~\ref{tab:LCDM}).

For selected examples, we also show 2D corner plots with marginalized 1D posterior probability density functions (PDF).
All results reported in what follows, including all relative errors and the constraints appearing in the figures, indicate $90\%$ C.I.~around the measured median value and are given for the AKK (AKS) waveform model.

\subsection{The GR limit}
\label{subsec:Xi=1}

\begin{table*}[htbp]
\centering
\renewcommand{\arraystretch}{1.5}
\caption{Median value and 90\% C.I.~of the marginalised posterior PDF of all cosmological parameters (from the fifth column), each obtained using two different waveform models (AKS and AKK), for an injected value $\Xi_0=1.0$ (for the injected values of the other parameters, see Sec.~\ref{sec:modGWpro}) and observation time of 4 years. Results are grouped according to: the considered EMRI model (first column), the cosmological scenario (second column, cf.~Sec.~\ref{sec:method}), the number of standard siren events used in the cosmological inference (third and fourth columns).}
\label{tab:10}
\resizebox{\textwidth}{!}{ 
\begin{tabular}{|c|c||c|c||c|c||c|c||c|c||c|c|}
\hline
\multirow{2}{*}{Model} & \multirow{2}{*}{Parameters}
& \multicolumn{2}{c||}{Num of StSi}
& \multicolumn{2}{c||}{$h$} 
& \multicolumn{2}{c||}{$\Omega_M$}  
& \multicolumn{2}{c||}{$\Xi_0$}
& \multicolumn{2}{c|}{$n$} \\
\cline{3-12}
 && AKS & AKK & AKS & AKK  & AKS & AKK & AKS & AKK & AKS & AKK \\
\hline \multirow{5}{*}{M1}
& $h+\Omega_M+\Xi_0+n$ & 7 & 12 & $0.704^{+0.049}_{-0.069}$ & $0.654^{+0.054}_{-0.045}$ & $0.278^{+0.198}_{-0.213}$ & $0.336^{+0.149}_{-0.241}$ & $1.016^{+0.714}_{-0.364}$ & $0.907^{+0.595}_{-0.272}$ & $1.008^{+1.743}_{-0.932}$ & $1.154^{+1.658}_{-1.085}$ \\ 
\cline{2-12}
& $h+\Omega_M+\Xi_0$ & 8 & 12 & $0.667^{+0.054}_{-0.046}$ & $0.679^{+0.036}_{-0.032}$ & $0.290^{+0.190}_{-0.222}$ & $0.330^{+0.155}_{-0.245}$ & $0.949^{+0.287}_{-0.242}$ & $1.020^{+0.183}_{-0.204}$ & - & - \\ 
\cline{2-12}
& $h+\Xi_0$ & 8 & 12 & $0.644^{+0.086}_{-0.036}$ & $0.656^{+0.060}_{-0.043}$ & - & - & $0.807^{+0.308}_{-0.159}$ & $0.930^{+0.170}_{-0.196}$ & - & - \\ 
\cline{2-12}
& $\Xi_0+n$ & 8 & 12 & - & - & - & - & $1.018^{+0.630}_{-0.273}$ & $1.022^{+0.633}_{-0.247}$ & $0.432^{+2.108}_{-0.411}$ & $0.323^{+2.150}_{-0.307}$ \\ \cline{2-12}
& $\Xi_0$ & 8 & 13 & - & - & - & - & $0.992^{+0.101}_{-0.067}$ & $0.985^{+0.031}_{-0.032}$ & - & - \\ 
\cline{2-12}
\hline \multirow{5}{*}{M5}
& $h+\Omega_M+\Xi_0+n$ & 1 & 2 & - & $0.691^{+0.062}_{-0.078}$ & - & $0.316^{+0.167}_{-0.241}$ & - & $0.970^{+0.793}_{-0.345}$ & - & $1.322^{+1.524}_{-1.230}$ \\ 
\cline{2-12}
& $h+\Omega_M+\Xi_0$ & 1 & 2 & - & $0.695^{+0.059}_{-0.084}$ & - & $0.292^{+0.189}_{-0.225}$ & - & $1.079^{+0.518}_{-0.428}$ & - & - \\ 
\cline{2-12}
& $h+\Xi_0$ & 1 & 2 & - & $0.696^{+0.057}_{-0.073}$ & - & - & - & $0.794^{+0.685}_{-0.165}$ & - & - \\ 
\cline{2-12}
& $\Xi_0+n$ & 1 & 2 & - & - & - & - & - & $0.838^{+0.870}_{-0.209}$ & - & $1.437^{+1.373}_{-1.390}$ \\ 
\cline{2-12}
& $\Xi_0$ & 1 & 2 & - & - & - & - & - & $1.011^{+0.197}_{-0.320}$ & - & - \\ 
\cline{2-12} \hline \multirow{5}{*}{M6}
& $h+\Omega_M+\Xi_0+n$ & 9 & 24 & $0.675^{+0.041}_{-0.038}$ & $0.670^{+0.019}_{-0.018}$ & $0.306^{+0.174}_{-0.233}$ & $0.328^{+0.154}_{-0.224}$ & $0.984^{+0.706}_{-0.329}$ & $1.027^{+0.625}_{-0.325}$ & $1.001^{+1.751}_{-0.928}$ & $0.815^{+1.916}_{-0.765}$ \\ 
\cline{2-12}
& $h+\Omega_M+\Xi_0$ & 9 & 26 & $0.664^{+0.055}_{-0.043}$ & $0.687^{+0.025}_{-0.024}$ & $0.285^{+0.193}_{-0.218}$ & $0.329^{+0.156}_{-0.237}$ & $0.917^{+0.286}_{-0.240}$ & $1.069^{+0.157}_{-0.232}$ & - & - \\ 
\cline{2-12}
& $h+\Xi_0$ & 9 & 27 & $0.688^{+0.066}_{-0.070}$ & $0.666^{+0.025}_{-0.024}$ & - & - & $0.949^{+0.585}_{-0.302}$ & $0.980^{+0.092}_{-0.090}$ & - & - \\ 
\cline{2-12}
& $\Xi_0+n$ & 9 & 28 & - & - & - & - & $1.033^{+0.654}_{-0.294}$ & $0.997^{+0.494}_{-0.255}$ & $0.492^{+2.064}_{-0.458}$ & $0.278^{+2.216}_{-0.268}$ \\ 
\cline{2-12}
& $\Xi_0$ & 10 & 29 & - & - & - & - & $0.967^{+0.089}_{-0.072}$ & $1.007^{+0.019}_{-0.018}$ & - & - \\ 
\cline{2-12} \hline
\end{tabular}}
\end{table*}

\begin{figure*}
     \includegraphics[width = 0.28\textwidth]{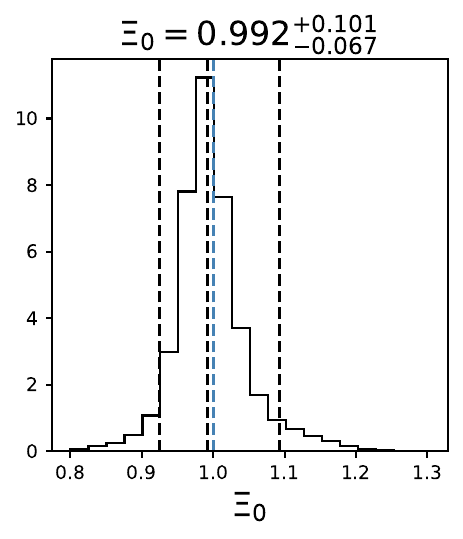}
     \includegraphics[width = 0.30\textwidth]{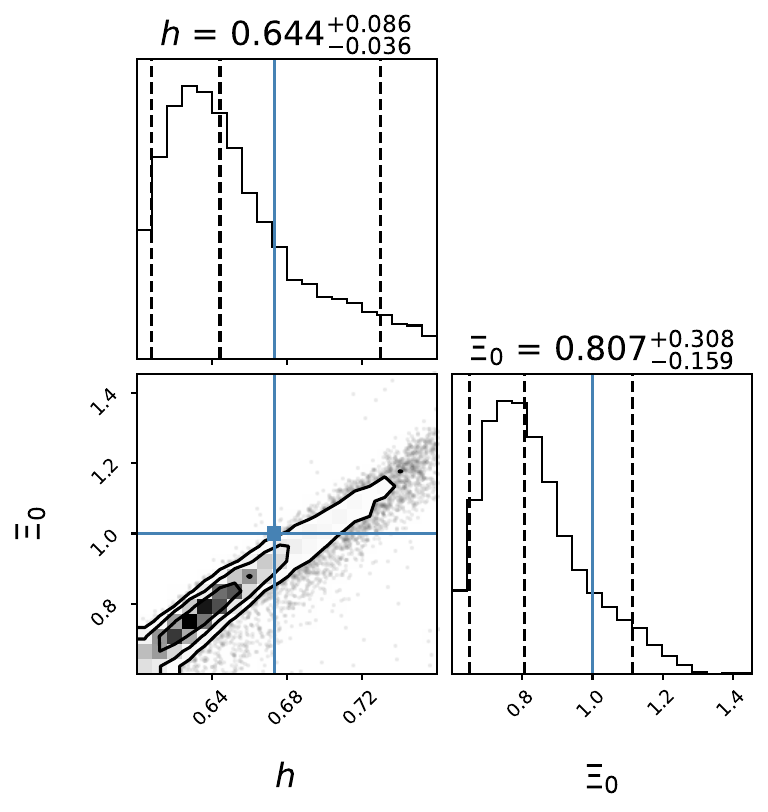}
     \includegraphics[width = 0.30\textwidth]{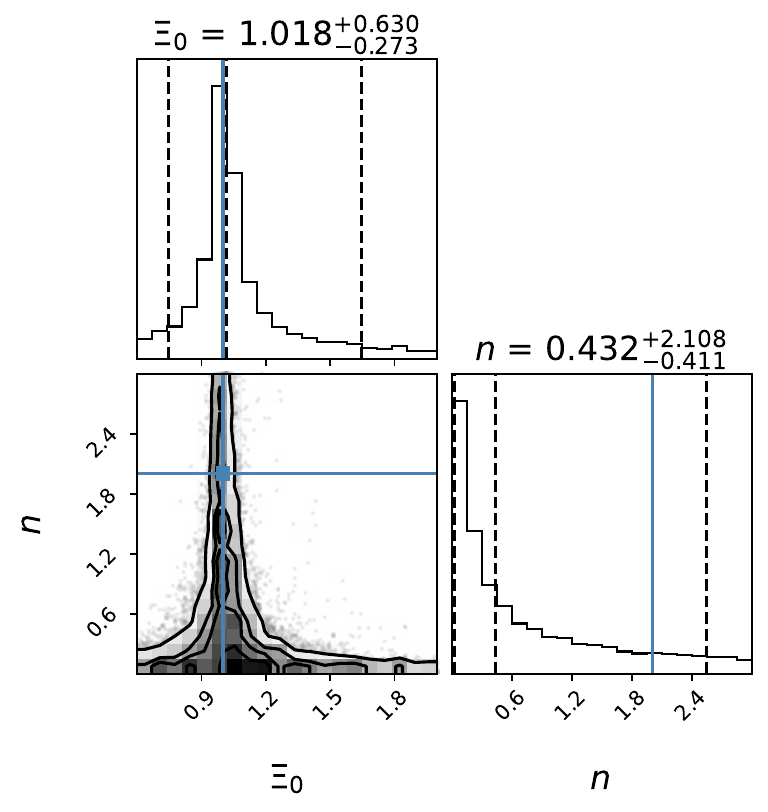}
     \includegraphics[width = 0.39\textwidth]{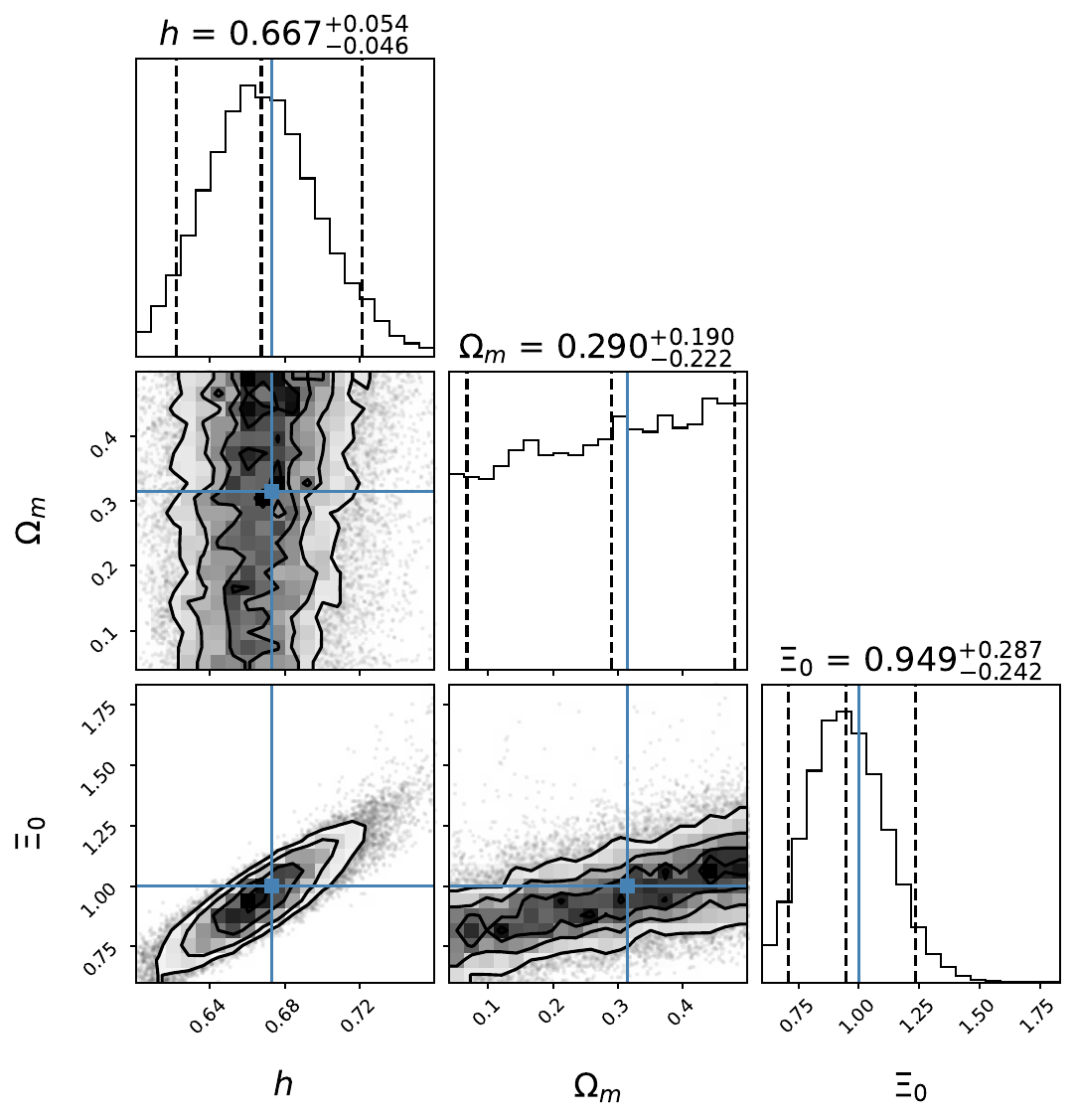}
     \includegraphics[width = 0.50\textwidth]{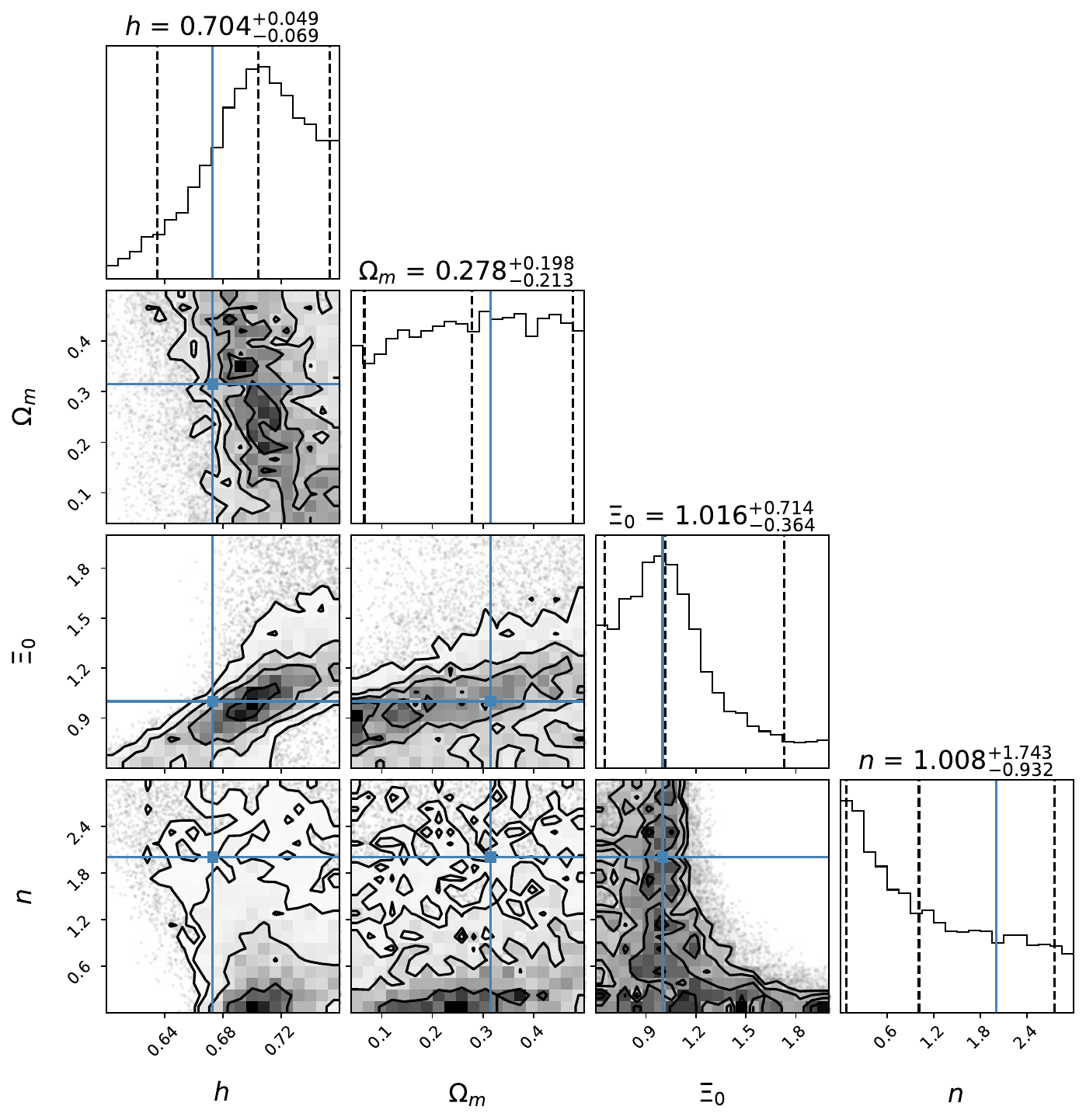}
     \caption{Corner plots (median and 90\% C.I., black dashed lines) of the five different parameter set scenarios for the M1 model and the AKS waveform. The injected value (blue line) for $\Xi_0$ is 1.}
     \label{fig:M1_4yr1.0}
\end{figure*}

Here we consider an injected value $\Xi_0=1.0$.
In Table~\ref{tab:10} we report the constraints from the 5 different modified gravity cosmological scenarios, for all EMRI population models (M1, M5, M6), and for both waveform models (AKK, AKS).
As an example, Fig.~\ref{fig:M1_4yr1.0} shows corner plots for these 5 scenarios in the M1 case with the AKS waveform.
All results assume 4 years of LISA observation (see Sec.~\ref{subsec:10yr} for the 10-year case).

First of all, we comment on the number of dark sirens that are used in each scenario.
As mentioned in Sec.~\ref{sec:method}, the actual number of EMRIs that end up being used in the cosmological inference depends on the cosmological scenario under study, since the boundaries of the error boxes in redshift space depend on the assumed prior on the cosmological parameters and consequently on the cosmology itself.
This limits to some extent our choice of the prior bounds, since choosing larger priors would increase the EMRI localisation error volumes in redshift space, resulting in more events that are ultimately not covered by our galaxy catalog and thus excluded from our analysis.
Since each parameter has the same prior in all cosmological scenarios, the number of dark sirens is the same regardless of the injected cosmological parameters, i.e., the number of dark sirens is the same as in the scenarios analysed in Sec.~\ref{subsec:Xi=0912} (but not Secs.~\ref{subsec:10yr} and \ref{subsec:LCDM}, where we assume a longer observation time and a different cosmological model, respectively).
For the M1 model, the number of available dark sirens in 4 years using the AKK (AKS) waveform model concentrates around 12 (8), while for M5 it is 2 (1) and for M6 it ranges from 24 (9) to 29 (10).

Assuming perfect knowledge of the $\Lambda$CDM parameters and only measuring the modified gravity parameter $\Xi_0$ (thus fixing $n$), EMRIs could yield a relative precision $\Delta\Xi_0/\Xi_0$ around $3.2\%$ ($8.5\%$) for M1, $26\%$ ($\gg 100\%$) for M5, and $1.8\%$ ($8.3\%$) for M6.
For comparison, on the theory side, the ``minimal'' RT non-local model predicts $\Xi_0\approx0.934$, a 6.6\% deviation from $\Xi_0$ = 1, with potentially larger values if initial conditions change~\cite{maggiore2017nonlocal,Belgacem:2019tbw}.
This is comparable to the results we obtain in the one-parameter cosmological scenario.
It should be noted that in this scenario we fix $h$ to its fiducial value, which may somehow limit the impact of the 1D constraint on $\Xi_0$, since from Eq.~\eqref{eq:Xi0} we can see that these parameters are correlated, which can lead to some degeneracy. It is therefore more interesting to perform a simultaneous inference on both parameters.

In the two-parameter scenario $h+\Xi_0$ instead, the relative precision $\Delta\Xi_0/\Xi_0$ drops to approximately $20\%$ ($29\%$) for M1, $54\%$ ($\gg 100\%$) for M5, and $9.3\%$ ($47\%$) for M6, while the relative accuracy $\Delta h/h$ is around $7.9\%$ ($9.5\%$) for M1, $9.3\%$ ($\gg 100\%$) for M5, and $3.7\%$ ($9.9\%$) for M6.
From Fig.~\ref{fig:M1_4yr1.0} we can moreover see that a strong correlation exists between $\Xi_0$ and $h$, as expected from Eq.~\eqref{eq:Xi0}.

In the cosmological scenario $\Xi_0+n$, where the two modified gravity parameters are allowed to vary, the relative precision $\Delta\Xi_0/\Xi_0$ becomes approximately $43\%$ ($44\%$) for M1, $68\%$ ($\gg 100\%$) for M5, $38\%$ ($46\%$) for M6, while $n$ cannot be measured in any of the scenarios.
We also note that the introduction of another modified gravity parameter $n$ causes the precision on $\Xi_0$ to drop drastically.

Next, we let the matter density parameter $\Omega_M$ vary and consider the three-parameter space $h+\Omega_M+\Xi_0$ and the four-parameter model $h+\Omega_M+\Xi_0+n$.
In the three parameter scenario, $\Omega_M$ can be constrained with a large relative accuracy of roughly 61\% (71\%) for M1, 71\% ($\gg$100\%) for M5, and 60\% (72\%) for M6.
These constraints basically do not change in the four-parameter model with the addition of $n$.
Introducing just $\Omega_M$ in general does not appreciably affect the measurements of $h$ and $\Xi_0$, which remain similar to the previous 2D scenarios.
On the other hand, introducing a secondary modified gravity parameter like $n$ worsens the constraints on $\Xi_0$, while it leaves the measurements of $h$ qualitatively unchanged.
In general, the constraints on $h$ see minimal changes between a two-, three-, and four-parameter cosmological scenarios.

Comparing the waveform models, in general the AKK model always provides better cosmological measurements than the AKS, since AKK yields more detections and generally provides larger SNRs.
Comparing between different population models, in models M1 and M6 the AKS waveform leads to comparable numbers of dark sirens, thus similar constraints.
M5 has only one event satisfying our selection criteria for the analysis,
thus we cannot obtain any interesting cosmological measurement.
Even using the AKK waveform, except for the 1D $\Xi_0$-case, we find that M5 tends to fail in recovering the true value or even providing a unimodal posterior: this is partly expected due to the lack of information in the data, which does not allow to break the correlations between $\Xi_0$ and the other parameters.
Regarding the results obtained with the AKK waveform, the accuracy on $h, \Omega_M, \Xi_0$ with M6 is always better, although often only slightly, than the corresponding results with M1, with the best measurement of $\Xi_0$ reaching $\sim$2\% in the $\Xi_0$-only scenario. On the other hand, the measurements with M5 are up to 10 times worse than with M1.
Finally, the parameter $n$ cannot be measured independently of the waveform used in any of the scenarios.

\subsection{Degeneracy between $\Xi_0$ and $n$}\label{subsec:Xi=0912}

\begin{figure*}
     \includegraphics[width = 0.30\textwidth]{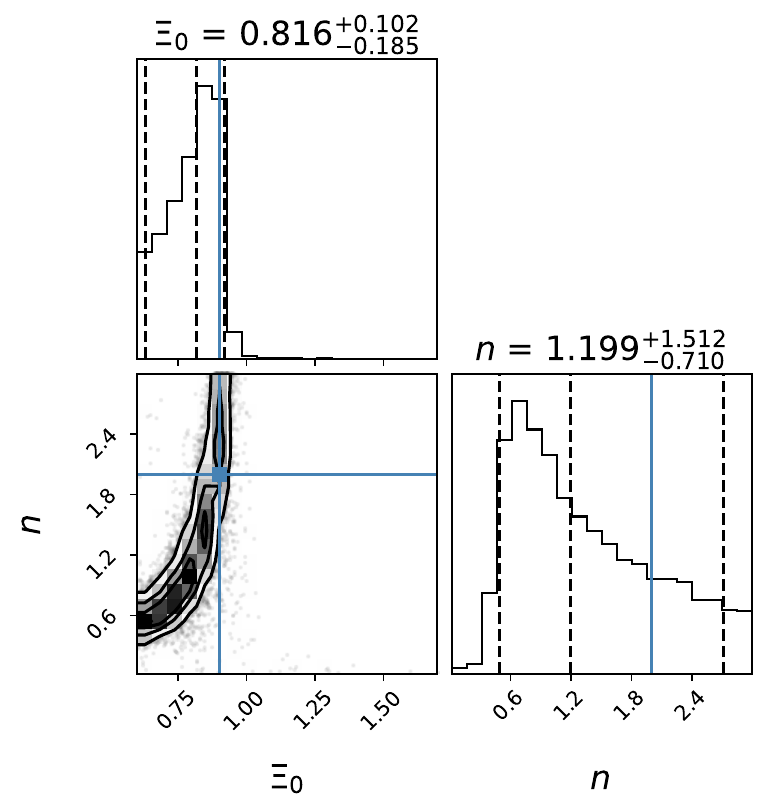}
     \includegraphics[width = 0.30\textwidth]{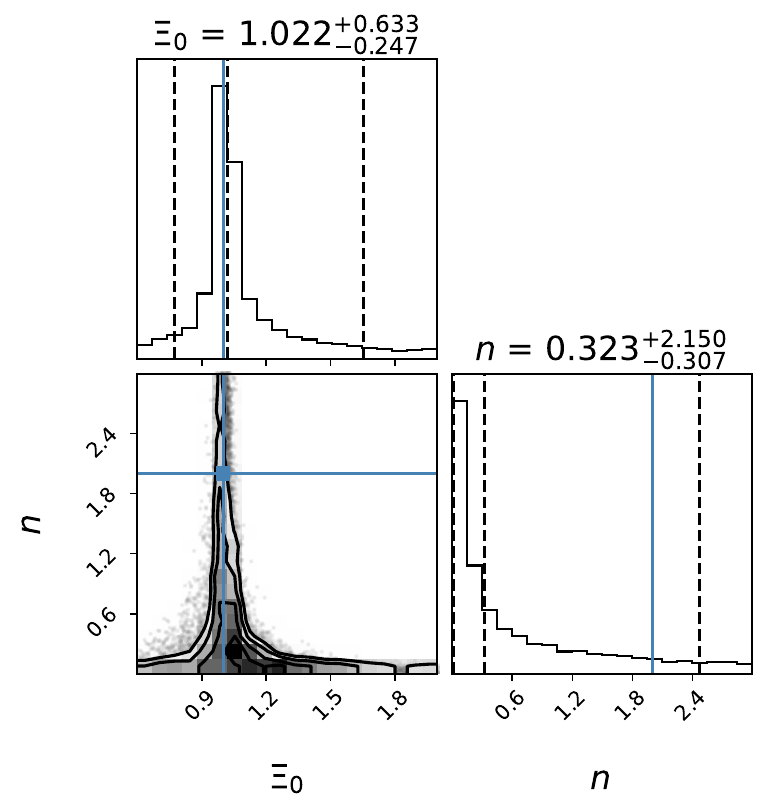}
     \includegraphics[width = 0.30\textwidth]{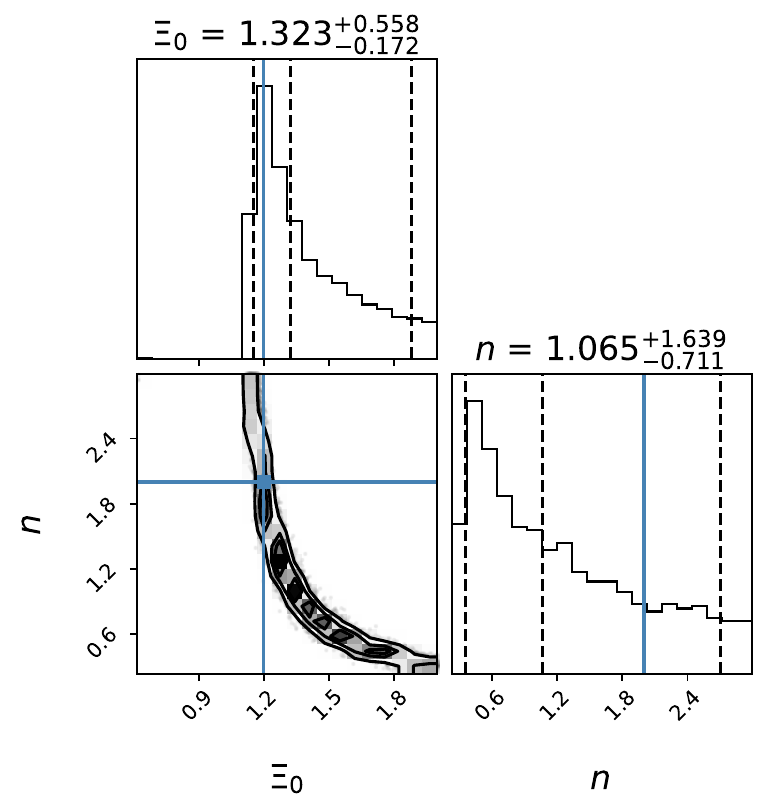}
     \caption{Corner plots (median and 90\% C.I., black dashed lines) of the $\Xi_0+n$ scenario for the M1 model, using the AKK waveform, showing the different correlation structures. From left to right, the injected value (in blue) for $\Xi_0$ is 0.9, 1.0, and 1.2.}
     \label{fig:M1_Xi0+n}
 \end{figure*}

Table~\ref{tab:09} and Table~\ref{tab:12} show the results for the injected values $\Xi_0 = 0.9$ and $\Xi_0 = 1.2$.
In these cases we only change the injected value of $\Xi_0$, keeping the injected values of all the other cosmological parameters the same.
We also stress that the number of standard sirens used in each scenario is equivalent to the $\Xi_0 = 1$ case, since it depends only on the cosmological priors used in the generation of the localisation error volumes, which is the same in all the cases; see Table~\ref{tab:10}.

We find that relative errors do not depend too much on the injected value of $\Xi_0$. Taking M1 as an example, in the one-parameter scenario the injected value of $\Xi_0=0.9$ leads to a relative uncertainty of 6.3\% (7.7\%), while the injected value $\Xi_0=1.2$ gives 2.9\% (8.3\%).
These numbers can be compared with the precision of 3.2\% (8.5\%) obtained in the case $\Xi_0=1$.
Some difference may be expected as different values of $\Xi_0$ change the luminosity distance-redshift relation (cf. Eq.~\eqref{eq:Xi0}) and consequently the parameter correlation structure, as discussed below. 
A similar trend can be found for all other higher-parameter scenarios; cf.~Tables \ref{tab:10}, \ref{tab:09}, and \ref{tab:12}.

It is interesting to analyse the 2-parameter $\Xi_0+n$ scenario in more detail.
We observe that although $n$ can never be measured, it has a strong degeneracy with $\Xi_0$.
This correlation is due to the low-redshift limit of the relationship given by Eq.~\eqref{eq:Xi0}: ${d_L^{\mathrm{GW}}(z)}/{d_L^{\mathrm{EM}}(z)}\approx1+n(\Xi_0-1)z$ as $z\rightarrow0$. 
At low redshift the two parameters are thus completely degenerate
along a straight line 
in the ($\Xi_0, n$) parameter space.
This leads to a strong degeneracy at intermediate redshift, which is clearly visible from Fig.~\ref{fig:M1_Xi0+n}.
The degeneracy has a qualitative different behaviour for $\Xi_0<1, \Xi_0=1, \Xi_0>1$, as it is evident from Fig.~\ref{fig:M1_Xi0+n}, which shows $\Xi_0+n$ corner plots for different fiducial values of $\Xi_0$.
Such a behaviour is directly related to the $n(\Xi_0-1)$ factor that governs the low-redshift limits, whose sign determines the $z \rightarrow 0$  degenerate line in the ($\Xi_0, n$) parameter space.
This degeneracy would be eventually resolved by adding higher redshift dark sirens to the cosmological inference.

\begin{table*}[htbp]
\centering
\renewcommand{\arraystretch}{1.5}
\caption{Median value and 90\% C.I.~of the marginalised posterior PDF of all cosmological parameters (from the third column), each obtained using two different waveform models (AKS and AKK), for an injected value $\Xi_0=0.9$ (for the injected values of the other parameters, see Sec.~\ref{sec:modGWpro}) and observation time of 4 years. Results are grouped according to: the considered EMRI model (first column), the cosmological scenario (second column, cf.~Sec.~\ref{sec:method}).
The number of dark sirens used in each scenario is the same as that shown in Table~\ref{tab:10}.}
\label{tab:09}
\resizebox{\textwidth}{!}{ 
\begin{tabular}{|c|c||c|c||c|c||c|c||c|c|}
\hline
\multirow{2}{*}{Model} & \multirow{2}{*}{Parameters}
& \multicolumn{2}{c||}{$h$} 
& \multicolumn{2}{c||}{$\Omega_M$}  
& \multicolumn{2}{c||}{$\Xi_0$}
& \multicolumn{2}{c|}{$n$} \\
\cline{3-10}
 & & AKS & AKK  & AKS & AKK & AKS & AKK & AKS & AKK \\
\hline \multirow{5}{*}{M1}
& $h+\Omega_M+\Xi_0+n$ & $0.667^{+0.050}_{-0.052}$ & $0.688^{+0.030}_{-0.038}$ & $0.333^{+0.152}_{-0.245}$ & $0.345^{+0.141}_{-0.247}$ & $0.829^{+0.729}_{-0.210}$ & $0.900^{+0.502}_{-0.256}$ & $1.533^{+1.315}_{-1.423}$ & $1.091^{+1.679}_{-1.021}$ \\ \cline{2-10}
& $h+\Omega_M+\Xi_0$  & $0.680^{+0.062}_{-0.065}$ & $0.658^{+0.047}_{-0.036}$ & $0.333^{+0.155}_{-0.254}$ & $0.314^{+0.168}_{-0.231}$ & $0.821^{+0.423}_{-0.209}$ & $0.835^{+0.205}_{-0.188}$ & - & - \\ \cline{2-10}
& $h+\Xi_0$ & $0.673^{+0.063}_{-0.043}$ & $0.653^{+0.056}_{-0.041}$ & - & - & $0.881^{+0.226}_{-0.222}$ & $0.842^{+0.174}_{-0.133}$ & - & - \\ \cline{2-10}
& $\Xi_0+n$ & - & - & - & - & $0.811^{+0.139}_{-0.197}$ & $0.816^{+0.102}_{-0.185}$ & $1.701^{+1.161}_{-1.365}$ & $1.199^{+1.512}_{-0.710}$ \\ \cline{2-10}
& $\Xi_0$ & - & - & - & - & $0.852^{+0.064}_{-0.068}$ & $0.941^{+0.057}_{-0.061}$ & - & - \\ \cline{2-10}
\hline \multirow{5}{*}{M5}
& $h+\Omega_M+\Xi_0+n$ & - & $0.702^{+0.052}_{-0.087}$ & - & $0.325^{+0.160}_{-0.244}$ & - & $0.859^{+0.931}_{-0.234}$ & - & $1.541^{+1.309}_{-1.417}$ \\ \cline{2-10}
& $h+\Omega_M+\Xi_0$ & - & $0.693^{+0.061}_{-0.082}$ & - & $0.318^{+0.165}_{-0.241}$ & - & $0.932^{+0.776}_{-0.312}$ & - & - \\ \cline{2-10}
& $h+\Xi_0$ & - & $0.701^{+0.054}_{-0.089}$ & - & - & - & $0.883^{+0.794}_{-0.253}$ & - & - \\ \cline{2-10}
& $\Xi_0+n$ & - & - & - & - & - & $0.836^{+0.563}_{-0.197}$ & - & $1.339^{+1.441}_{-0.834}$ \\ \cline{2-10}
& $\Xi_0$ & - & - & - & - & - & $0.705^{+0.600}_{-0.093}$ & - & - \\ \cline{2-10}
\hline \multirow{5}{*}{M6}
& $h+\Omega_M+\Xi_0+n$ & $0.685^{+0.041}_{-0.044}$ & $0.672^{+0.022}_{-0.025}$ & $0.323^{+0.160}_{-0.242}$ & $0.367^{+0.122}_{-0.250}$ & $0.894^{+0.683}_{-0.250}$ & $0.862^{+0.342}_{-0.223}$ & $1.165^{+1.616}_{-1.086}$ & $1.100^{+1.643}_{-1.023}$ \\ \cline{2-10}
& $h+\Omega_M+\Xi_0$ & $0.657^{+0.057}_{-0.038}$ & $0.683^{+0.021}_{-0.021}$ & $0.300^{+0.182}_{-0.226}$ & $0.327^{+0.157}_{-0.237}$ & $0.820^{+0.316}_{-0.189}$ & $0.958^{+0.130}_{-0.188}$ & - & - \\ \cline{2-10}
& $h+\Xi_0$ & $0.702^{+0.050}_{-0.065}$ & $0.673^{+0.019}_{-0.018}$ & - & - & $1.042^{+0.216}_{-0.287}$ & $0.894^{+0.063}_{-0.059}$ & - & - \\ \cline{2-10}
& $\Xi_0+n$ & - & - & - & - & $0.835^{+0.105}_{-0.197}$ & $0.842^{+0.078}_{-0.204}$ & $1.224^{+1.526}_{-0.822}$ & $1.103^{+1.608}_{-0.675}$ \\ \cline{2-10}
& $\Xi_0$ & - & - & - & - & $0.902^{+0.076}_{-0.073}$ & $0.905^{+0.023}_{-0.021}$ & - & - \\ \cline{2-10}
\hline
\end{tabular}}
\end{table*}

\begin{table*}[htbp]
\centering
\renewcommand{\arraystretch}{1.5}
\caption{The same as in Table~\ref{tab:09}, but for an injected value $\Xi_0=1.2$ and observation time of 4 years.}
\label{tab:12}
\resizebox{\textwidth}{!}{ 
\begin{tabular}{|c|c||c|c||c|c||c|c||c|c|}
\hline
\multirow{2}{*}{Model} & \multirow{2}{*}{Parameters}
& \multicolumn{2}{c||}{$h$} 
& \multicolumn{2}{c||}{$\Omega_M$}  
& \multicolumn{2}{c||}{$\Xi_0$}
& \multicolumn{2}{c|}{$n$} \\
\cline{3-10}
 & & AKS & AKK  & AKS & AKK & AKS & AKK & AKS & AKK \\
\hline \multirow{5}{*}{M1}
& $h+\Omega_M+\Xi_0+n$ & $0.665^{+0.052}_{-0.039}$ & $0.663^{+0.040}_{-0.029}$ & $0.265^{+0.211}_{-0.201}$ & $0.264^{+0.211}_{-0.195}$ & $1.233^{+0.601}_{-0.493}$ & $1.227^{+0.593}_{-0.360}$ & $1.076^{+1.688}_{-0.966}$ & $0.954^{+1.787}_{-0.860}$ \\ \cline{2-10}
& $h+\Omega_M+\Xi_0$ & $0.666^{+0.074}_{-0.054}$ & $0.676^{+0.053}_{-0.045}$ & $0.293^{+0.188}_{-0.225}$ & $0.274^{+0.202}_{-0.209}$ & $1.093^{+0.362}_{-0.415}$ & $1.230^{+0.280}_{-0.248}$ & - & - \\ \cline{2-10}
& $h+\Xi_0$ & $0.656^{+0.073}_{-0.048}$ & $0.655^{+0.047}_{-0.051}$ & - & - & $1.025^{+0.335}_{-0.379}$ & $1.148^{+0.185}_{-0.526}$ & - & - \\ \cline{2-10}
& $\Xi_0+n$ & - & - & - & - & $1.295^{+0.566}_{-0.208}$ & $1.323^{+0.558}_{-0.172}$ & $0.985^{+1.695}_{-0.727}$ & $1.065^{+1.639}_{-0.711}$ \\ \cline{2-10}
& $\Xi_0$ & - & - & - & - & $1.093^{+0.079}_{-0.103}$ & $1.200^{+0.034}_{-0.035}$ & - & - \\ \cline{2-10}
\hline \multirow{5}{*}{M5}
& $h+\Omega_M+\Xi_0+n$ & - & $0.676^{+0.075}_{-0.065}$ & - & $0.265^{+0.210}_{-0.203}$ & - & $1.300^{+0.589}_{-0.575}$ & - & $1.192^{+1.577}_{-1.057}$ \\ \cline{2-10}
& $h+\Omega_M+\Xi_0$ & - & $0.695^{+0.059}_{-0.083}$ & - & $0.314^{+0.169}_{-0.239}$ & - & $0.856^{+0.755}_{-0.237}$ & - & - \\ \cline{2-10}
& $h+\Xi_0$ & - & $0.703^{+0.053}_{-0.090}$ & - & - & - & $0.955^{+0.532}_{-0.333}$ & - & - \\ \cline{2-10}
& $\Xi_0+n$ & - & - & - & - & - & $0.865^{+0.807}_{-0.240}$ & - & $1.344^{+1.483}_{-1.147}$ \\ \cline{2-10}
& $\Xi_0$ & - & - & - & - & - & $1.139^{+0.123}_{-0.273}$ & - & - \\ \cline{2-10}
\hline \multirow{5}{*}{M6}
& $h+\Omega_M+\Xi_0+n$ & $0.690^{+0.059}_{-0.058}$ & $0.654^{+0.022}_{-0.018}$ & $0.238^{+0.233}_{-0.179}$ & $0.270^{+0.207}_{-0.196}$ & $1.418^{+0.498}_{-0.597}$ & $1.209^{+0.599}_{-0.352}$ & $1.322^{+1.463}_{-1.138}$ & $0.909^{+1.852}_{-0.819}$ \\ \cline{2-10}
& $h+\Omega_M+\Xi_0$ & $0.693^{+0.050}_{-0.063}$ & $0.666^{+0.025}_{-0.024}$ & $0.275^{+0.205}_{-0.213}$ & $0.321^{+0.162}_{-0.236}$ & $1.228^{+0.313}_{-0.405}$ & $1.171^{+0.165}_{-0.211}$ & - & - \\ \cline{2-10}
& $h+\Xi_0$ & $0.654^{+0.071}_{-0.045}$ & $0.677^{+0.028}_{-0.024}$ & - & - & $1.092^{+0.363}_{-0.250}$ & $1.225^{+0.112}_{-0.098}$ & - & - \\ \cline{2-10}
& $\Xi_0+n$ & - & - & - & - & $1.348^{+0.533}_{-0.228}$ & $1.232^{+0.554}_{-0.083}$ & $1.194^{+1.550}_{-0.843}$ & $1.606^{+1.253}_{-1.202}$ \\ \cline{2-10}
& $\Xi_0$ & - & - & - & - & $1.125^{+0.098}_{-0.219}$ & $1.202^{+0.022}_{-0.023}$ & - & - \\ \cline{2-10}
\hline
\end{tabular}}
\end{table*}

\subsection{LISA observation time: 4 vs 10 yr}
\label{subsec:10yr}

In this subsection, we discuss constraints obtained with an extension of LISA science observation up to 10 years. 
We assume an injected value $\Xi_0=1$ and present all results in Table~\ref{tab:10yr10}
Similar results apply as well to the analysis of other non-GR values, which therefore we will not discuss in detail here.
Note that the number of dark sirens used for the 10-year cosmological analyses is, by construction, $10/4$ greater than in the 4-year case (cf.~Table~\ref{tab:10}).

\begin{figure*}
    \centering
    \begin{tabular}{ccc}
        \vspace{0.2cm}
           & \textbf{4 years, AKK Model} &   \\
         Model M1 & Model M5 & Model M6 \\
        \includegraphics[width = 0.32\textwidth]{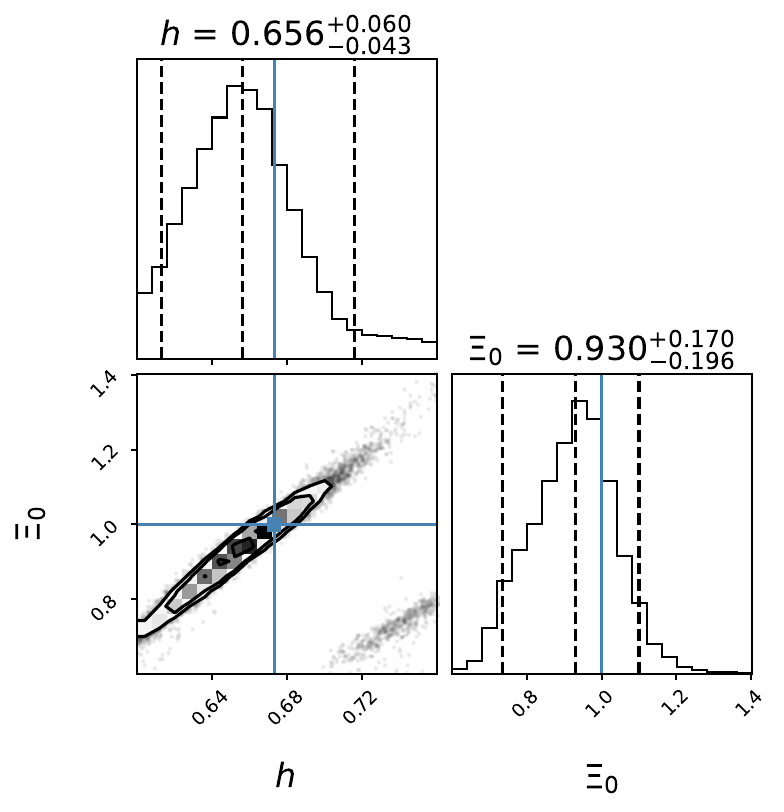}&
        \includegraphics[width = 0.32\textwidth]{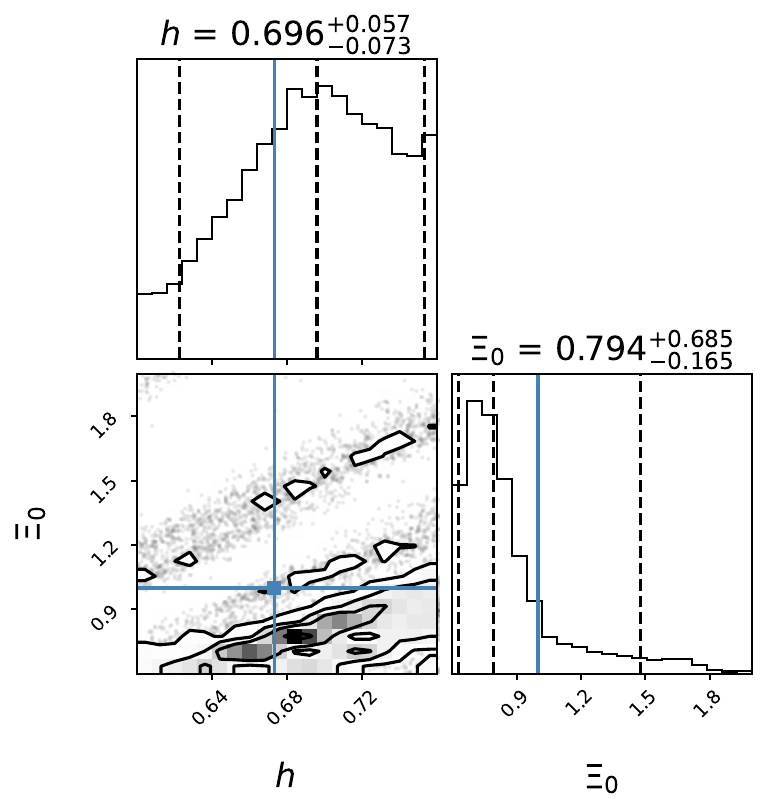}&
        \includegraphics[width = 0.32\textwidth]{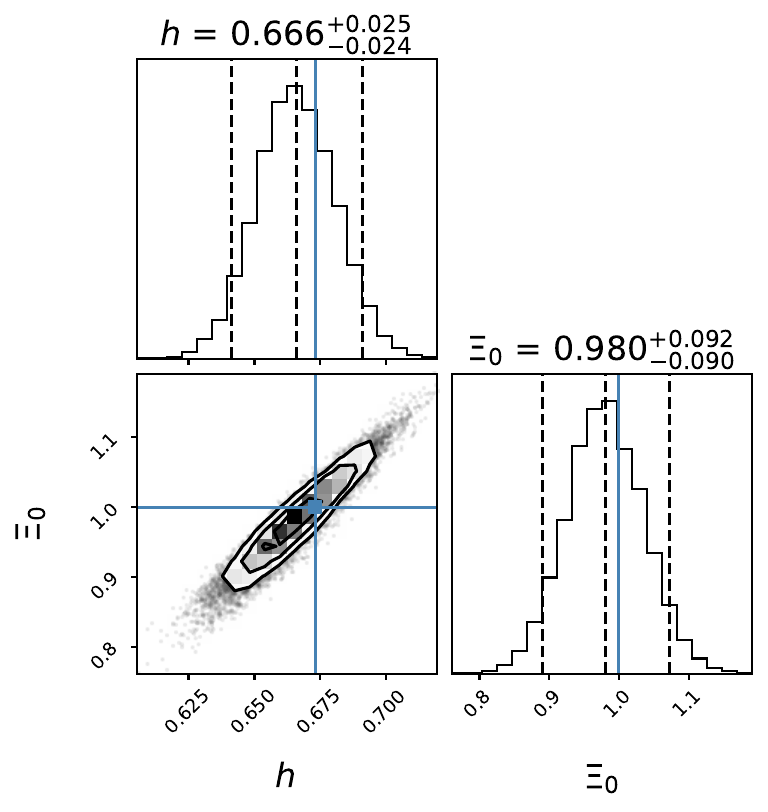}\\
        \vspace{0.2cm}
           & \textbf{10 years, AKK Model} &   \\
         Model M1 & Model M5 & Model M6 \\
        \includegraphics[width = 0.32\textwidth]{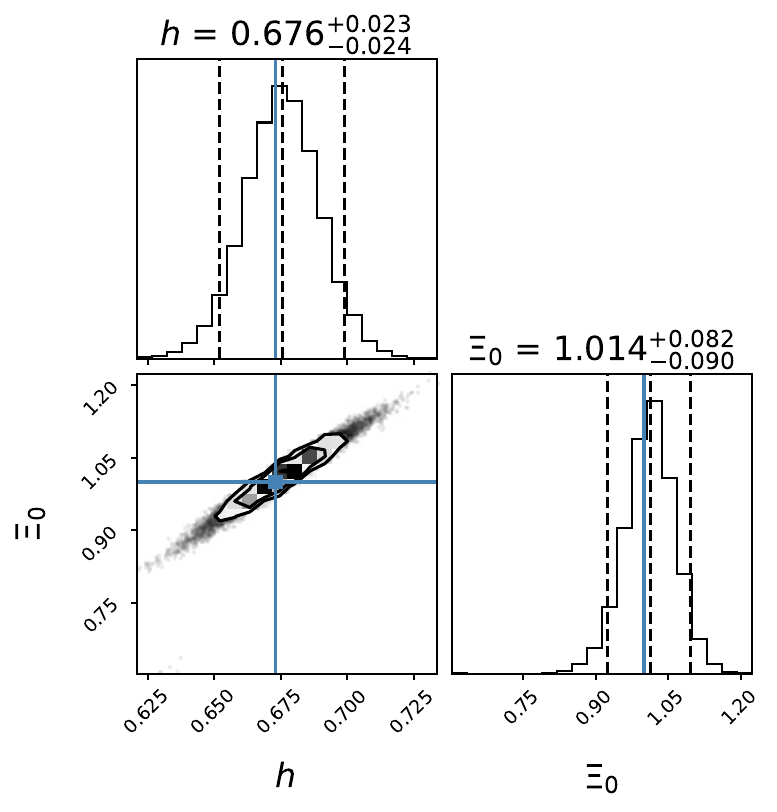}&
        \includegraphics[width = 0.32\textwidth]{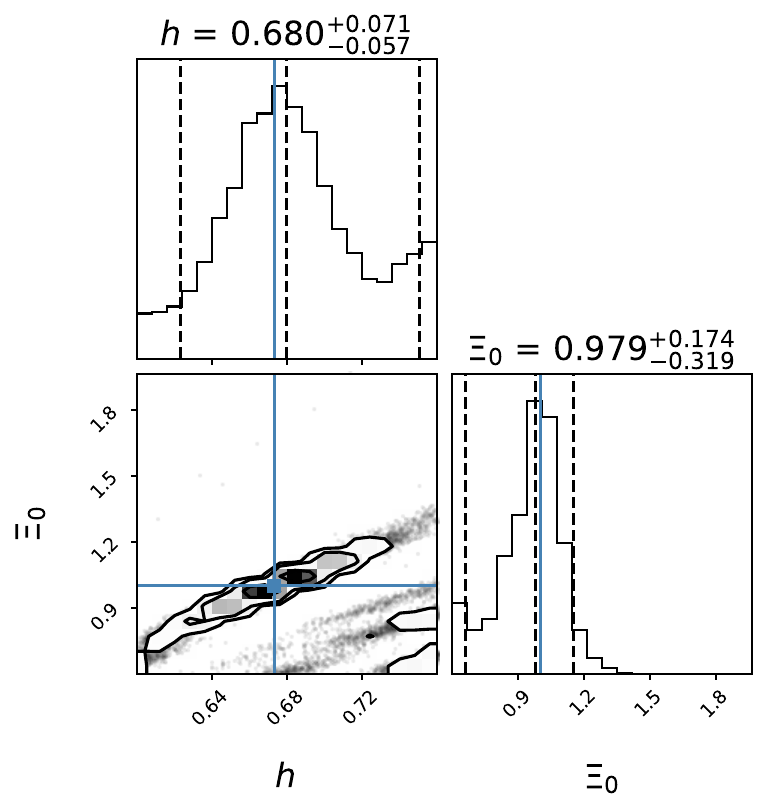}&
        \includegraphics[width = 0.32\textwidth]{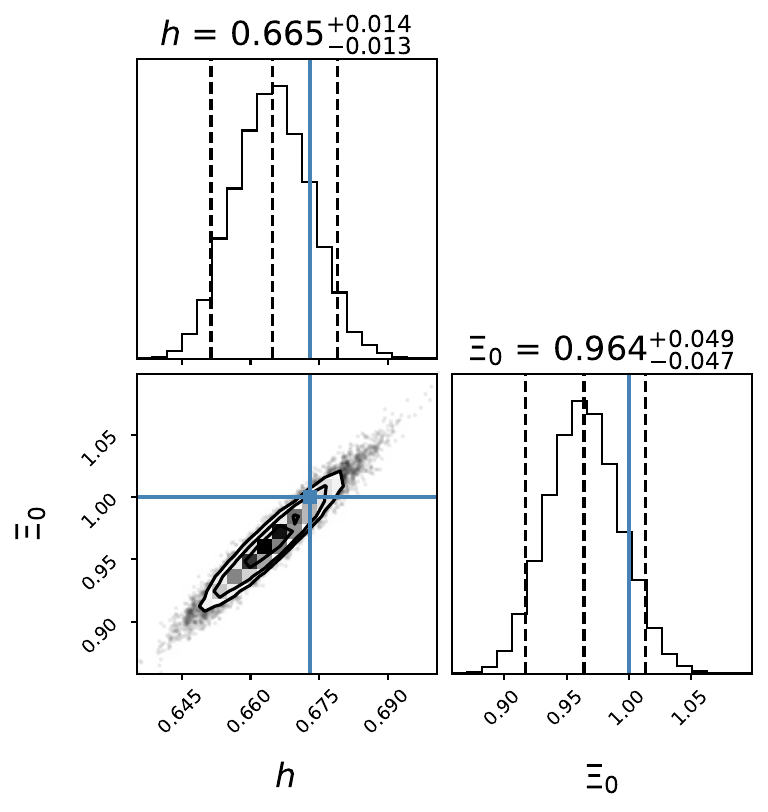}\\
    \end{tabular}
    \caption{From left to right, corner plots (median and 90\% C.I., black dashed lines) for the $h+\Xi_0$ two-parameter scenario using the AKK model for the M1, M5, and M6 models, respectively. The LISA observation time of the upper (lower) panels is 4 years (10 years). The injected value (in blue) of $\Xi_0$ is 1.}
    \label{fig:M1M5M6_h+Xi0}
\end{figure*}

In the one-parameter scenario we find $\Delta\Xi_0/\Xi_0 \sim$ 1.8\% (4.0\%) for M1, 6.5\% (42\%) for M5, and 1.2\% (4.3\%) for M6.
For the two-parameter $h+\Xi_0$ scenario we find instead $\Delta\Xi_0/\Xi_0 \sim$ 8.5\% (18\%) for M1, 25\% (41\%) for M5, and 5.0\% (18\%) for M6, together with $\Delta h/h \sim$ 3.5\% (5.8\%) for M1, 9.4\% (10.4\%) for M5, and 2.0\% (5.0\%) for M6.
Fig.~\ref{fig:M1M5M6_h+Xi0} compares the results of the $h+\Xi_0$ scenario between 4 and 10 years of observation.
A strong correlation between $h$ and $\Xi_0$ is clearly shown, as we already noted in Sec.~\ref{subsec:Xi=1}, revealing the intrinsic connection between the two parameters which control the first derivative of $d^{\rm GW}_L(z)$ as $z\rightarrow0$, namely the Hubble law for GWs.

Comparing Table~\ref{tab:10yr10} with Table~\ref{tab:10}, the precision for 10 years of observation is on average 1.4 times better than the one obtained with 4 years of observation, which is slightly better than, but in rough accordance with, $\sqrt{N_{\rm 10yr}/N_{\rm 4yr}} = 1.58$.
However, the improvement in each scenario varies greatly due to the low statistics of detected dark sirens, and no consistent trend can be found for M5 due to the extremely low number of events used for the cosmological inference, especially in 4 years of observations.
We also note that only 10 years of observations allow M5 to provide interesting constraints on the cosmological parameters.

\begin{table*}[htbp]
\centering
\renewcommand{\arraystretch}{1.5}
\caption{Median value and 90\% C.I.~of the marginalised posterior PDF of all cosmological parameters (from the fifth column), each obtained using two different waveform models (AKS and AKK), for an injected value $\Xi_0=1.0$ (for the injected values of the other parameters, see Sec.~\ref{sec:modGWpro}) and observation time of 10 years. Results are grouped according to: the considered EMRI model (first column), the cosmological scenario (second column, cf.~Sec.~\ref{sec:method}), the number of standard siren events used in the cosmological inference (third and fourth columns).}
\label{tab:10yr10}
\resizebox{\textwidth}{!}{ 
\begin{tabular}{|c|c||c|c||c|c||c|c||c|c||c|c|}
\hline
\multirow{2}{*}{Model} & \multirow{2}{*}{Parameters}
& \multicolumn{2}{c||}{Num of StSi}
& \multicolumn{2}{c||}{$h$} 
& \multicolumn{2}{c||}{$\Omega_M$}  
& \multicolumn{2}{c||}{$\Xi_0$}
& \multicolumn{2}{c|}{$n$} \\
\cline{3-12}
 && AKS & AKK & AKS & AKK  & AKS & AKK & AKS & AKK & AKS & AKK \\
\hline \multirow{5}{*}{M1}
& $h+\Omega_M+\Xi_0+n$ & 17 & 29 & $0.660^{+0.026}_{-0.032}$ & $0.664^{+0.019}_{-0.020}$ & $0.329^{+0.155}_{-0.245}$ & $0.342^{+0.143}_{-0.249}$ & $0.941^{+0.651}_{-0.281}$ & $0.957^{+0.564}_{-0.287}$ & $0.960^{+1.772}_{-0.898}$ & $0.869^{+1.829}_{-0.821}$ \\ \cline{2-12}
& $h+\Omega_M+\Xi_0$ & 19 & 30 & $0.681^{+0.037}_{-0.031}$ & $0.659^{+0.030}_{-0.025}$ & $0.292^{+0.189}_{-0.223}$ & $0.332^{+0.152}_{-0.245}$ & $0.995^{+0.197}_{-0.209}$ & $0.941^{+0.155}_{-0.174}$ & - & - \\ \cline{2-12}
& $h+\Xi_0$ & 19 & 30 & $0.647^{+0.042}_{-0.033}$ & $0.676^{+0.023}_{-0.024}$ & - & - & $0.876^{+0.169}_{-0.148}$ & $1.014^{+0.082}_{-0.090}$ & - & - \\ \cline{2-12}
& $\Xi_0+n$ & 19 & 32 & - & - & - & - & $0.975^{+0.438}_{-0.273}$ & $1.024^{+0.598}_{-0.182}$ & $0.422^{+2.092}_{-0.403}$ & $0.302^{+2.187}_{-0.288}$ \\ \cline{2-12}
& $\Xi_0$ & 19 & 33 & - & - & - & - & $0.987^{+0.038}_{-0.041}$ & $1.004^{+0.019}_{-0.018}$ & - & - \\ \cline{2-12}
\hline \multirow{5}{*}{M5}
& $h+\Omega_M+\Xi_0+n$ & 3 & 5 & $0.722^{+0.035}_{-0.091}$ & $0.681^{+0.058}_{-0.061}$ & $0.335^{+0.151}_{-0.254}$ & $0.314^{+0.169}_{-0.231}$ & $0.834^{+0.766}_{-0.220}$ & $1.015^{+0.697}_{-0.351}$ & $1.586^{+1.309}_{-1.482}$ & $1.069^{+1.691}_{-0.992}$ \\ \cline{2-12}
& $h+\Omega_M+\Xi_0$ & 3 & 6 & $0.697^{+0.058}_{-0.085}$ & $0.667^{+0.065}_{-0.057}$ & $0.301^{+0.181}_{-0.232}$ & $0.305^{+0.177}_{-0.232}$ & $0.853^{+0.738}_{-0.230}$ & $0.941^{+0.289}_{-0.258}$ & - & - \\ \cline{2-12}
& $h+\Xi_0$ & 3 & 6 & $0.693^{+0.061}_{-0.083}$ & $0.680^{+0.071}_{-0.057}$ & - & - & $0.880^{+0.470}_{-0.249}$ & $0.979^{+0.174}_{-0.319}$ & - & - \\ \cline{2-12}
& $\Xi_0+n$ & 3 & 6 & - & - & - & - & $0.983^{+0.670}_{-0.315}$ & $1.000^{+0.517}_{-0.250}$ & $0.665^{+1.981}_{-0.635}$ & $0.328^{+2.180}_{-0.316}$ \\ \cline{2-12}
& $\Xi_0$ & 3 & 7 & - & - & - & - & $1.004^{+0.505}_{-0.330}$ & $1.010^{+0.028}_{-0.104}$ & - & - \\ \cline{2-12}
\hline \multirow{5}{*}{M6}
& $h+\Omega_M+\Xi_0+n$ & 23 & 60 & $0.671^{+0.033}_{-0.030}$ & $0.669^{+0.013}_{-0.014}$ & $0.327^{+0.157}_{-0.242}$ & $0.351^{+0.132}_{-0.223}$ & $0.923^{+0.667}_{-0.268}$ & $1.011^{+0.603}_{-0.323}$ & $1.036^{+1.693}_{-0.966}$ & $0.725^{+1.941}_{-0.674}$ \\ \cline{2-12}
& $h+\Omega_M+\Xi_0$ & 23 & 65 & $0.667^{+0.034}_{-0.029}$ & $0.669^{+0.025}_{-0.020}$ & $0.298^{+0.183}_{-0.226}$ & $0.359^{+0.130}_{-0.250}$ & $0.931^{+0.216}_{-0.196}$ & $0.988^{+0.145}_{-0.162}$ & - & - \\ \cline{2-12}
& $h+\Xi_0$ & 23 & 68 & $0.660^{+0.035}_{-0.031}$ & $0.665^{+0.014}_{-0.013}$ & - & - & $0.951^{+0.163}_{-0.184}$ & $0.964^{+0.049}_{-0.047}$ & - & - \\ \cline{2-12}
& $\Xi_0+n$ & 23 & 69 & - & - & - & - & $1.008^{+0.618}_{-0.262}$ & $0.993^{+0.383}_{-0.240}$ & $0.358^{+2.138}_{-0.341}$ & $0.265^{+2.191}_{-0.256}$ \\ \cline{2-12}
& $\Xi_0$ & 24 & 72 & - & - & - & - & $0.963^{+0.042}_{-0.041}$ & $1.001^{+0.013}_{-0.012}$ & - & - \\ \cline{2-12}
\hline
\end{tabular}}
\end{table*}

\subsection{$\Lambda$CDM}
\label{subsec:LCDM}

Compared to the EMRI cosmological study presented in \cite{Laghi:2021pqk}, in this study we improved some aspects of the analysis method as well as the computation of the likelihood; see Sec.~\ref{sec:method} for details.
It is therefore interesting to verify the impact of these changes in the forecasts for the classic $\Lambda$CDM  model compared to~\cite{Laghi:2021pqk}.
The main differences in the method can be summarized as follows: (1) In the construction of the error boxes, we use 3$\sigma$ errors on distance and sky localization measurements in order to find the boundaries in redshift space. In contrast, the study in \cite{Laghi:2021pqk} relied on 2$\sigma$ errors on distance and sky localization; (2) We use an updated galaxy light cone produced with a state-of-the-art semi-analytic evolution model~\cite{villalba19}, while \cite{Laghi:2021pqk} used an older full sky map from the Virgo Millennium Database~\cite{VMD}; (3) The code used in this study implements an optimised numerical marginalisation over the GW redshift, as discussed in Sec.~\ref{sec:method}; (4) Here we assume a uniform prior ranges on $h$ of [0.6, 0.76], while \cite{Laghi:2021pqk} assumed [0.6, 0.86]; (5) In this study we report results for the median realisation over the five realisations analysed, whereas \cite{Laghi:2021pqk} reports results averaged over three realisations; (6) In \cite{Laghi:2021pqk} only results for the AKK waveform were available, whereas here we also include results for the AKS model.

Table~\ref{tab:LCDM} shows our results for $\Lambda$CDM, where we simultaneously infer the values of $h$ and $\Omega_M$.
We find that constraints on both $h$ and $\Omega_M$ improve by a factor of 1.6 at most, although they are often indistinguishable from those provided in \cite{Laghi:2021pqk}.
For 4-year observations with the AKK waveform, the 90\% C.I. constraints now are $\sim$ 0.021 (3.1\%), 0.044 (6.4\%), 0.015 (2.2\%) on $h$ and $\sim$ 0.140 (36\%), 0.207 (64\%), 0.093 (28\%) on $\Omega_M$ for M1, M5, M6, respectively.
We also provide results with the AKS waveform for the first time, which, as expected, yield worse limits than the AKK waveform due to the lower number of detected dark sirens.

\begin{table*}[htbp]
\centering
\renewcommand{\arraystretch}{1.5}
\caption{
Median value and 90\% C.I.~of the marginalised posterior PDF of the $\Lambda$CDM cosmological parameters (from the fifth column), each obtained using two different waveform models (AKS and AKK), for both observation times of 4 and 10 years. Results are grouped according to: the considered EMRI model (first column), the observation time (second column), the number of standard siren events used in the cosmological inference (third and fourth columns).}
\label{tab:LCDM}
\begin{tabular}{|c|c||c|c||c|c||c|c|}
\hline
\multirow{2}{*}{Model} & \multirow{2}{*}{Observation}
& \multicolumn{2}{c||}{Num of StSi}
& \multicolumn{2}{c||}{$h$} 
& \multicolumn{2}{c|}{$\Omega_M$}   \\
\cline{3-8}
 && AKS & AKK & AKS & AKK  & AKS & AKK  \\
\hline \multirow{2}{*}{M1}
& 4yr & 8 & 13 & $0.675^{+0.035}_{-0.022}$ & $0.661^{+0.027}_{-0.016}$ & $0.345^{+0.139}_{-0.219}$ & $0.392^{+0.094}_{-0.185}$ \\ \cline{2-8}
& 10yr & 20 & 33 & $0.674^{+0.022}_{-0.018}$ & $0.665^{+0.016}_{-0.014}$ & $0.309^{+0.161}_{-0.185}$ & $0.376^{+0.096}_{-0.101}$ \\ \cline{2-8}
\hline \multirow{2}{*}{M5}
& 4yr & 1 & 3 & - & $0.686^{+0.045}_{-0.042}$ & - & $0.325^{+0.162}_{-0.252}$ \\ \cline{2-8}
& 10yr & 3 & 7 & $0.721^{+0.035}_{-0.083}$ & $0.667^{+0.029}_{-0.020}$ & $0.299^{+0.181}_{-0.222}$ & $0.353^{+0.133}_{-0.167}$ \\ \cline{2-8}
\hline \multirow{2}{*}{M6}
& 4yr & 9 & 31 & $0.677^{+0.023}_{-0.022}$ & $0.674^{+0.014}_{-0.015}$ & $0.283^{+0.198}_{-0.221}$ & $0.330^{+0.104}_{-0.081}$ \\ \cline{2-8}
& 10yr & 23 & 77 & $0.668^{+0.020}_{-0.014}$ & $0.672^{+0.009}_{-0.010}$ & $0.343^{+0.136}_{-0.185}$ & $0.325^{+0.063}_{-0.051}$ \\ \cline{2-8}
\hline
\end{tabular}
\end{table*}


\section{Discussion and conclusion}
\label{sec:conclusion}

In this section we discuss our findings and compare them with the literature. 

We found that over 4 years of LISA observations, constraints on $\Xi_0$ can reach a few percent level (90\% C.I.) in our fiducial (M1) and optimistic (M6) scenarios, assuming all other parameters are fixed.
By inferring simultaneously other parameters, especially $n$, the accuracy on $\Xi_0$ drops quickly.
We have shown that EMRIs cannot be used to perform full inference on both $\Lambda$CDM and modified gravity parameters, unless some of them are constrained by narrow priors.
Our estimates are improved by considering an extended LISA observation period of 10 years, in which case $\Xi_0$ is constrained at the level of a few percent overall, even reaching almost 1\% (90\% C.I.) in the most optimistic scenarios.
These results are equally degraded when additional parameters are inferred together with $\Xi_0$.

Current LVK observations show no evidence for deviations from GR at cosmological scales \cite{abbott2019tests,abbott2021tests}, though constraints on an anomalous damping in the propagation of GWs are weak.
To compare our results with them, here we first summarize the investigations focusing on constraints and predictions using LVK data. 
Ref.~\cite{Finke:2021aom} uses O1 to O3a data to constrain the parameter $\Xi_0$, fixing all other parameters, employing a dark siren analysis correlating with the galaxy catalog GLADE~\cite{dalya2018glade}.
Their 68\% C.I.~result reads $\Xi_0 = 2.1^{+3.2}_{-1.2}$.
Ref.~\cite{Mancarella:2021ecn} instead target black hole (BH) binaries  from GWTC-3~\cite{LIGOScientific:2021djp} using them as spectral sirens to jointly constrain 4 population inference parameters with $\Xi_0$.
They give $\Xi_0 = 1.2^{+0.7}_{-0.7}$ with a flat prior, and $\Xi_0 = 1.2^{+0.4}_{-0.8}$ with a uniform prior in log (68\% C.I.).
The same authors provide also forecasts with 5 years of LIGO/Virgo data at design sensitivity, estimating that the accuracy may range from 10\% (for an injected $\Xi_0=1$) to 20\% (for $\Xi_0=$ 1.8).
On the other hand, Ref.~\cite{Mastrogiovanni:2020mvm} use the events GW170817 and GW190521 as bright sirens (assuming both EM events have a confident associated EM counterpart), and different waveforms to constrain $h, \Omega_M, \Xi_0$ and $n$.
Assuming a Planck (flat) prior on $h$ they give an upper limit on $\Xi_0<10\, (4.4)$, but no limit on $\Omega_M$ and $n$.
Ref.~\cite{Leyde:2022orh} performed a Bayesian spectral siren analysis (thus without using any galaxy catalog) including an estimation on population parameters.
They also perform a model selection based on different population and modified GW propagation models.
They estimate that $\Xi_0$ can be constrained at the 51\% level with $\sim 90$ binary BH detections at O4-like sensitivity, and at the 20\% level with an additional $\sim 400$ binary BH detections at O5-like sensitivity. 
Finally, a recent study from Ref.~\cite{chen2023testing} analysed 45 events (42 binary BHs and 3 neutron star-black holes) from GWTC-3 as dark standard sirens and the GLADE+ galaxy catalog~\cite{dalya2022glade+} to infer jointly $H_0$, $\Xi_0$, $n$ and the GW population parameters, finding (68\% C.I.) $\Xi_0=1.67^{+0.93}_{-0.94}$ and $n=0.80^{+3.58}_{-0.69}$.
In terms of forecasts with LVK detectors at design sensitivity, \cite{mukherjee2021testing} perform a cross-correlation of galaxy surveys with dark sirens and additional information from the baryon acoustic oscillation scale to constrain $\Xi_0$ marginalising over redshift dependence, forecasting $\Xi_0=0.98^{+0.04}_{-0.23}$ with $\sim3500$ binary BH dark sirens distributed up to redshift $z\sim 0.5$ detected by the LVK network at design sensitivity.
The LISA constraints on $\Xi_0$ from EMRIs that we report here are in general few times better than results obtained with current LVK results, and possibly even better than future forecasts for current (second generation) GW detectors, at least for realistic and optimistic scenarios.
We stress however that unlike most of the results from LVK analyses mentioned above, our results do not take into account the population parameters of the EMRIs.

On the other hand, our results are comparable to expectations from 3G detectors only in the most optimistic cases.
Ref.~\cite{Belgacem:2019tbw} estimates constraints on $\Xi_0, w_0$, together with $\Lambda$CDM parameters, using data from CMB, BAO, SNe, as well as standard siren data from simulated binary neutron star mergers and their corresponding GRB observations from a network of 3G detectors. 
They report that the accuracy on $\Xi_0$ could reach below percent levels with 10 years of data (with a 80\% duty cycle for each detector).
Ref.~\cite{branchesi2023science} produced different estimate measurements of $\Xi_0+n$ using different designs of the Einstein Telescope (ET)~\cite{hild2008pushing,punturo2010einstein,hild2011sensitivity}, the future European ground-based observatory designed to detect GW sources up to high redshift.
Without EM counterparts, but using information on the tidal deformability of neutron stars, the accuracy on $\Xi_0$ fluctuates between 3\% to 10\%, depending on the geometry and sensitivity of ET. 
Using a similar approach, Ref.~\cite{Jiang:2021mpd} further investigate constraints on $\Xi_0$ with binary neutron star mergers observed by both ET and DECIGO~\cite{kawamura2011japanese,kawamura2021current}.
FM constraints on $\Xi_0$ using ET (ET+DECIGO) are at the level of $\sim$ 3\% (1\%) with an observation time of 3 years, assuming part of the events will be associated to an EM counterpart.
In addition, Ref.~\cite{Finke:2021znb} show that with $\sim 6$ quadruply-lensed GW events per year detected by ET, in 4 years $\Xi_0$ could be detected at the $\sim6\%$ level.
3G detectors should thus confidently constrain $\Xi_0$ to a few percent level or better (see also~\cite{matos2021gravitational}), making them more effective at measuring deviations in the GW propagation than EMRI observations with LISA.

However, other GW sources can be used as standard sirens, like those that will be detected by LISA, and thus provide further cosmological constraints.
Ref.~\cite{LISACosmologyWorkingGroup:2019mwx} uses LISA massive black hole binaries (MBHBs) with an EM counterpart as bright sirens, showing that $\Xi_0$ can be measured at the 1\% to 4.4\% accuracy (68\% C.I.), depending on different MBHB population models.
Furthermore, Ref~\cite{Yang:2021qge} show that combining data from future GW bright sirens detected by ground-based detectors, LISA, and Taiji, and adding current Planck+BAO+Pantheon data, could constrain $\Xi_0$ at the level of 0.46\% (68\% C.I.).
Such results suggest that the combination of EMRI data with other LISA cosmological measurements (or even with other space-based GW observations) could yield constraints on $\Xi_0$ comparable to 3G forecasts.
In our results, the limits on $\Xi_0$ alone from the fiducial model M1 and optimistic model M6 using the AKK waveform are already comparable to the ones derived by LISA MBHB bright sirens, implying that combining the two datasets could further improve these bounds.
Moreover, since EMRIs are mostly distributed at low redshift and MBHBs are detected at higher redshift, their cosmological results will complement each other, possibly breaking some of the degeneracies in the parameter space of some scenarios \cite{Tamanini:2016uin}.
Additionally many theoretical models of DE predicts a deviation from $\Lambda$CDM for both modified GW propagation parameters and DE equation of state parameters, meaning that in they should be inferred simultaneously for realistic DE scenarios~\cite{Belgacem:2018lbp}.

In conclusion, we find that the forecasts that we can place with LISA EMRIs as dark sirens, assuming a galaxy catalogs complete up to $z=1$, are definitely better than what LVK data can provide now and in the future, but comparable to 3G forecasts only in the most optimistic cases.
The investigation we provide here is the first dark siren analysis on beyond-GR deviations with LISA EMRI data, and it helps to better define the cosmological science case of LISA.
Further work will be necessary to fully comprehend the cosmological potential of the LISA mission, especially by combining data from different populations of LISA standard sirens, in particular EMRI and MBHBs.

\section*{Acknowledgements}

The authors would like to thank Stas Babak, Enrico Barausse, and Alberto Sesana for providing the original EMRI catalogs from \cite{Babak:2017tow}, David Izquierdo-Villalba for providing the galaxy catalog used in the dark siren analysis, and Niccolò Muttoni for help with the implementation of the cross-correlation between GW events and galaxies. C. L. also thanks Lijing Shao, Ziming Wang, and Dicong Liang for helpful discussions. 
The authors acknowledge support form the French space agency CNES in the framework of LISA.
This project has received financial support from the CNRS through the MITI interdisciplinary programs. C. L. is supported by the China Scholarship Council (CSC) and the National Natural Science Foundation of China (11991053, 11975027). D.L. acknowledges funding for this work from the CNES Postdoctoral Fellowship program.

\bibliographystyle{apsrev}
\bibliography{refs} 

\begin{thebibliography}{77}
\expandafter\ifx\csname natexlab\endcsname\relax\def\natexlab#1{#1}\fi
\expandafter\ifx\csname bibnamefont\endcsname\relax
  \def\bibnamefont#1{#1}\fi
\expandafter\ifx\csname bibfnamefont\endcsname\relax
  \def\bibfnamefont#1{#1}\fi
\expandafter\ifx\csname citenamefont\endcsname\relax
  \def\citenamefont#1{#1}\fi
\expandafter\ifx\csname url\endcsname\relax
  \def\url#1{\texttt{#1}}\fi
\expandafter\ifx\csname urlprefix\endcsname\relax\def\urlprefix{URL }\fi
\providecommand{\bibinfo}[2]{#2}
\providecommand{\eprint}[2][]{\url{#2}}

\bibitem[{\citenamefont{Abdalla et~al.}(2022)}]{Abdalla:2022yfr}
\bibinfo{author}{\bibfnamefont{E.}~\bibnamefont{Abdalla}} \bibnamefont{et~al.},
  \bibinfo{journal}{JHEAp} \textbf{\bibinfo{volume}{34}}, \bibinfo{pages}{49}
  (\bibinfo{year}{2022}), \eprint{2203.06142}.

\bibitem[{\citenamefont{Huterer and Shafer}(2018)}]{Huterer:2017buf}
\bibinfo{author}{\bibfnamefont{D.}~\bibnamefont{Huterer}} \bibnamefont{and}
  \bibinfo{author}{\bibfnamefont{D.~L.} \bibnamefont{Shafer}},
  \bibinfo{journal}{Rept. Prog. Phys.} \textbf{\bibinfo{volume}{81}},
  \bibinfo{pages}{016901} (\bibinfo{year}{2018}), \eprint{1709.01091}.

\bibitem[{\citenamefont{Schutz}(1986)}]{Schutz:1986gp}
\bibinfo{author}{\bibfnamefont{B.~F.} \bibnamefont{Schutz}},
  \bibinfo{journal}{Nature} \textbf{\bibinfo{volume}{323}},
  \bibinfo{pages}{310} (\bibinfo{year}{1986}).

\bibitem[{\citenamefont{Holz and Hughes}(2005)}]{Holz:2005df}
\bibinfo{author}{\bibfnamefont{D.~E.} \bibnamefont{Holz}} \bibnamefont{and}
  \bibinfo{author}{\bibfnamefont{S.~A.} \bibnamefont{Hughes}},
  \bibinfo{journal}{Astrophys. J.} \textbf{\bibinfo{volume}{629}},
  \bibinfo{pages}{15} (\bibinfo{year}{2005}), \eprint{astro-ph/0504616}.

\bibitem[{\citenamefont{Abbott
  et~al.}(2017{\natexlab{a}})}]{LIGOScientific:2017adf}
\bibinfo{author}{\bibfnamefont{B.~P.} \bibnamefont{Abbott}}
  \bibnamefont{et~al.} (\bibinfo{collaboration}{LIGO Scientific, Virgo, 1M2H,
  Dark Energy Camera GW-E, DES, DLT40, Las Cumbres Observatory, VINROUGE,
  MASTER}), \bibinfo{journal}{Nature} \textbf{\bibinfo{volume}{551}},
  \bibinfo{pages}{85} (\bibinfo{year}{2017}{\natexlab{a}}),
  \eprint{1710.05835}.

\bibitem[{\citenamefont{Soares-Santos et~al.}(2019)\citenamefont{Soares-Santos,
  Palmese, Hartley, Annis, Garcia-Bellido, Lahav, Doctor, Fishbach, Holz, Lin
  et~al.}}]{soares2019first}
\bibinfo{author}{\bibfnamefont{M.}~\bibnamefont{Soares-Santos}},
  \bibinfo{author}{\bibfnamefont{A.}~\bibnamefont{Palmese}},
  \bibinfo{author}{\bibfnamefont{W.}~\bibnamefont{Hartley}},
  \bibinfo{author}{\bibfnamefont{J.}~\bibnamefont{Annis}},
  \bibinfo{author}{\bibfnamefont{J.}~\bibnamefont{Garcia-Bellido}},
  \bibinfo{author}{\bibfnamefont{O.}~\bibnamefont{Lahav}},
  \bibinfo{author}{\bibfnamefont{Z.}~\bibnamefont{Doctor}},
  \bibinfo{author}{\bibfnamefont{M.}~\bibnamefont{Fishbach}},
  \bibinfo{author}{\bibfnamefont{D.}~\bibnamefont{Holz}},
  \bibinfo{author}{\bibfnamefont{H.}~\bibnamefont{Lin}}, \bibnamefont{et~al.},
  \bibinfo{journal}{The Astrophysical Journal Letters}
  \textbf{\bibinfo{volume}{876}}, \bibinfo{pages}{L7} (\bibinfo{year}{2019}).

\bibitem[{\citenamefont{Gray et~al.}(2020)\citenamefont{Gray, Hernandez, Qi,
  Sur, Brady, Chen, Farr, Fishbach, Gair, Ghosh et~al.}}]{gray2020cosmological}
\bibinfo{author}{\bibfnamefont{R.}~\bibnamefont{Gray}},
  \bibinfo{author}{\bibfnamefont{I.~M.} \bibnamefont{Hernandez}},
  \bibinfo{author}{\bibfnamefont{H.}~\bibnamefont{Qi}},
  \bibinfo{author}{\bibfnamefont{A.}~\bibnamefont{Sur}},
  \bibinfo{author}{\bibfnamefont{P.~R.} \bibnamefont{Brady}},
  \bibinfo{author}{\bibfnamefont{H.-Y.} \bibnamefont{Chen}},
  \bibinfo{author}{\bibfnamefont{W.~M.} \bibnamefont{Farr}},
  \bibinfo{author}{\bibfnamefont{M.}~\bibnamefont{Fishbach}},
  \bibinfo{author}{\bibfnamefont{J.~R.} \bibnamefont{Gair}},
  \bibinfo{author}{\bibfnamefont{A.}~\bibnamefont{Ghosh}},
  \bibnamefont{et~al.}, \bibinfo{journal}{Physical Review D}
  \textbf{\bibinfo{volume}{101}}, \bibinfo{pages}{122001}
  (\bibinfo{year}{2020}).

\bibitem[{\citenamefont{Abbott et~al.}(2023)}]{LIGOScientific:2021aug}
\bibinfo{author}{\bibfnamefont{R.}~\bibnamefont{Abbott}} \bibnamefont{et~al.}
  (\bibinfo{collaboration}{LIGO Scientific, Virgo,, KAGRA, VIRGO}),
  \bibinfo{journal}{Astrophys. J.} \textbf{\bibinfo{volume}{949}},
  \bibinfo{pages}{76} (\bibinfo{year}{2023}), \eprint{2111.03604}.

\bibitem[{\citenamefont{Mastrogiovanni
  et~al.}(2021{\natexlab{a}})\citenamefont{Mastrogiovanni, Leyde, Karathanasis,
  Chassande-Mottin, Steer, Gair, Ghosh, Gray, Mukherjee, and
  Rinaldi}}]{Mastrogiovanni:2021wsd}
\bibinfo{author}{\bibfnamefont{S.}~\bibnamefont{Mastrogiovanni}},
  \bibinfo{author}{\bibfnamefont{K.}~\bibnamefont{Leyde}},
  \bibinfo{author}{\bibfnamefont{C.}~\bibnamefont{Karathanasis}},
  \bibinfo{author}{\bibfnamefont{E.}~\bibnamefont{Chassande-Mottin}},
  \bibinfo{author}{\bibfnamefont{D.~A.} \bibnamefont{Steer}},
  \bibinfo{author}{\bibfnamefont{J.}~\bibnamefont{Gair}},
  \bibinfo{author}{\bibfnamefont{A.}~\bibnamefont{Ghosh}},
  \bibinfo{author}{\bibfnamefont{R.}~\bibnamefont{Gray}},
  \bibinfo{author}{\bibfnamefont{S.}~\bibnamefont{Mukherjee}},
  \bibnamefont{and} \bibinfo{author}{\bibfnamefont{S.}~\bibnamefont{Rinaldi}},
  \bibinfo{journal}{Phys. Rev. D} \textbf{\bibinfo{volume}{104}},
  \bibinfo{pages}{062009} (\bibinfo{year}{2021}{\natexlab{a}}),
  \eprint{2103.14663}.

\bibitem[{\citenamefont{Gair et~al.}(2023)}]{Gair:2022zsa}
\bibinfo{author}{\bibfnamefont{J.~R.} \bibnamefont{Gair}} \bibnamefont{et~al.},
  \bibinfo{journal}{Astron. J.} \textbf{\bibinfo{volume}{166}},
  \bibinfo{pages}{22} (\bibinfo{year}{2023}), \eprint{2212.08694}.

\bibitem[{\citenamefont{Mastrogiovanni
  et~al.}(2023)\citenamefont{Mastrogiovanni, Laghi, Gray, Santoro, Ghosh,
  Karathanasis, Leyde, Steer, Perri\`es, and Pierra}}]{PhysRevD.108.042002}
\bibinfo{author}{\bibfnamefont{S.}~\bibnamefont{Mastrogiovanni}},
  \bibinfo{author}{\bibfnamefont{D.}~\bibnamefont{Laghi}},
  \bibinfo{author}{\bibfnamefont{R.}~\bibnamefont{Gray}},
  \bibinfo{author}{\bibfnamefont{G.~C.} \bibnamefont{Santoro}},
  \bibinfo{author}{\bibfnamefont{A.}~\bibnamefont{Ghosh}},
  \bibinfo{author}{\bibfnamefont{C.}~\bibnamefont{Karathanasis}},
  \bibinfo{author}{\bibfnamefont{K.}~\bibnamefont{Leyde}},
  \bibinfo{author}{\bibfnamefont{D.~A.} \bibnamefont{Steer}},
  \bibinfo{author}{\bibfnamefont{S.}~\bibnamefont{Perri\`es}},
  \bibnamefont{and} \bibinfo{author}{\bibfnamefont{G.}~\bibnamefont{Pierra}},
  \bibinfo{journal}{Phys. Rev. D} \textbf{\bibinfo{volume}{108}},
  \bibinfo{pages}{042002} (\bibinfo{year}{2023}).

\bibitem[{\citenamefont{Gray et~al.}(2023)\citenamefont{Gray, Beirnaert,
  Karathanasis, Revenu, Turski, Chen, Baker, Vallejo, Romano, Ghosh
  et~al.}}]{gray2023joint}
\bibinfo{author}{\bibfnamefont{R.}~\bibnamefont{Gray}},
  \bibinfo{author}{\bibfnamefont{F.}~\bibnamefont{Beirnaert}},
  \bibinfo{author}{\bibfnamefont{C.}~\bibnamefont{Karathanasis}},
  \bibinfo{author}{\bibfnamefont{B.}~\bibnamefont{Revenu}},
  \bibinfo{author}{\bibfnamefont{C.}~\bibnamefont{Turski}},
  \bibinfo{author}{\bibfnamefont{A.}~\bibnamefont{Chen}},
  \bibinfo{author}{\bibfnamefont{T.}~\bibnamefont{Baker}},
  \bibinfo{author}{\bibfnamefont{S.}~\bibnamefont{Vallejo}},
  \bibinfo{author}{\bibfnamefont{A.~E.} \bibnamefont{Romano}},
  \bibinfo{author}{\bibfnamefont{T.}~\bibnamefont{Ghosh}},
  \bibnamefont{et~al.}, \bibinfo{journal}{arXiv preprint arXiv:2308.02281}
  (\bibinfo{year}{2023}).

\bibitem[{\citenamefont{Amaro-Seoane et~al.}(2007)\citenamefont{Amaro-Seoane,
  Gair, Freitag, Miller, Mandel, Cutler, and Babak}}]{amaro2007intermediate}
\bibinfo{author}{\bibfnamefont{P.}~\bibnamefont{Amaro-Seoane}},
  \bibinfo{author}{\bibfnamefont{J.~R.} \bibnamefont{Gair}},
  \bibinfo{author}{\bibfnamefont{M.}~\bibnamefont{Freitag}},
  \bibinfo{author}{\bibfnamefont{M.~C.} \bibnamefont{Miller}},
  \bibinfo{author}{\bibfnamefont{I.}~\bibnamefont{Mandel}},
  \bibinfo{author}{\bibfnamefont{C.~J.} \bibnamefont{Cutler}},
  \bibnamefont{and} \bibinfo{author}{\bibfnamefont{S.}~\bibnamefont{Babak}},
  \bibinfo{journal}{Classical and Quantum Gravity}
  \textbf{\bibinfo{volume}{24}}, \bibinfo{pages}{R113} (\bibinfo{year}{2007}).

\bibitem[{\citenamefont{Babak et~al.}(2017)\citenamefont{Babak, Gair, Sesana,
  Barausse, Sopuerta, Berry, Berti, Amaro-Seoane, Petiteau, and
  Klein}}]{Babak:2017tow}
\bibinfo{author}{\bibfnamefont{S.}~\bibnamefont{Babak}},
  \bibinfo{author}{\bibfnamefont{J.}~\bibnamefont{Gair}},
  \bibinfo{author}{\bibfnamefont{A.}~\bibnamefont{Sesana}},
  \bibinfo{author}{\bibfnamefont{E.}~\bibnamefont{Barausse}},
  \bibinfo{author}{\bibfnamefont{C.~F.} \bibnamefont{Sopuerta}},
  \bibinfo{author}{\bibfnamefont{C.~P.~L.} \bibnamefont{Berry}},
  \bibinfo{author}{\bibfnamefont{E.}~\bibnamefont{Berti}},
  \bibinfo{author}{\bibfnamefont{P.}~\bibnamefont{Amaro-Seoane}},
  \bibinfo{author}{\bibfnamefont{A.}~\bibnamefont{Petiteau}}, \bibnamefont{and}
  \bibinfo{author}{\bibfnamefont{A.}~\bibnamefont{Klein}},
  \bibinfo{journal}{Phys. Rev. D} \textbf{\bibinfo{volume}{95}},
  \bibinfo{pages}{103012} (\bibinfo{year}{2017}), \eprint{1703.09722}.

\bibitem[{\citenamefont{Amaro-Seoane et~al.}(2017)}]{LISA:2017pwj}
\bibinfo{author}{\bibfnamefont{P.}~\bibnamefont{Amaro-Seoane}}
  \bibnamefont{et~al.} (\bibinfo{collaboration}{LISA}) (\bibinfo{year}{2017}),
  \eprint{1702.00786}.

\bibitem[{\citenamefont{MacLeod and Hogan}(2008)}]{macleod2008precision}
\bibinfo{author}{\bibfnamefont{C.~L.} \bibnamefont{MacLeod}} \bibnamefont{and}
  \bibinfo{author}{\bibfnamefont{C.~J.} \bibnamefont{Hogan}},
  \bibinfo{journal}{Physical Review D} \textbf{\bibinfo{volume}{77}},
  \bibinfo{pages}{043512} (\bibinfo{year}{2008}).

\bibitem[{\citenamefont{Laghi et~al.}(2021)\citenamefont{Laghi, Tamanini,
  Del~Pozzo, Sesana, Gair, Babak, and Izquierdo-Villalba}}]{Laghi:2021pqk}
\bibinfo{author}{\bibfnamefont{D.}~\bibnamefont{Laghi}},
  \bibinfo{author}{\bibfnamefont{N.}~\bibnamefont{Tamanini}},
  \bibinfo{author}{\bibfnamefont{W.}~\bibnamefont{Del~Pozzo}},
  \bibinfo{author}{\bibfnamefont{A.}~\bibnamefont{Sesana}},
  \bibinfo{author}{\bibfnamefont{J.}~\bibnamefont{Gair}},
  \bibinfo{author}{\bibfnamefont{S.}~\bibnamefont{Babak}}, \bibnamefont{and}
  \bibinfo{author}{\bibfnamefont{D.}~\bibnamefont{Izquierdo-Villalba}},
  \bibinfo{journal}{Mon. Not. Roy. Astron. Soc.}
  \textbf{\bibinfo{volume}{508}}, \bibinfo{pages}{4512} (\bibinfo{year}{2021}),
  \eprint{2102.01708}.

\bibitem[{\citenamefont{Tamanini et~al.}(2016)\citenamefont{Tamanini, Caprini,
  Barausse, Sesana, Klein, and Petiteau}}]{Tamanini:2016zlh}
\bibinfo{author}{\bibfnamefont{N.}~\bibnamefont{Tamanini}},
  \bibinfo{author}{\bibfnamefont{C.}~\bibnamefont{Caprini}},
  \bibinfo{author}{\bibfnamefont{E.}~\bibnamefont{Barausse}},
  \bibinfo{author}{\bibfnamefont{A.}~\bibnamefont{Sesana}},
  \bibinfo{author}{\bibfnamefont{A.}~\bibnamefont{Klein}}, \bibnamefont{and}
  \bibinfo{author}{\bibfnamefont{A.}~\bibnamefont{Petiteau}},
  \bibinfo{journal}{JCAP} \textbf{\bibinfo{volume}{04}}, \bibinfo{pages}{002}
  (\bibinfo{year}{2016}), \eprint{1601.07112}.

\bibitem[{\citenamefont{Corman et~al.}(2022)\citenamefont{Corman, Ghosh,
  Escamilla-Rivera, Hendry, Marsat, and Tamanini}}]{Corman:2021avn}
\bibinfo{author}{\bibfnamefont{M.}~\bibnamefont{Corman}},
  \bibinfo{author}{\bibfnamefont{A.}~\bibnamefont{Ghosh}},
  \bibinfo{author}{\bibfnamefont{C.}~\bibnamefont{Escamilla-Rivera}},
  \bibinfo{author}{\bibfnamefont{M.~A.} \bibnamefont{Hendry}},
  \bibinfo{author}{\bibfnamefont{S.}~\bibnamefont{Marsat}}, \bibnamefont{and}
  \bibinfo{author}{\bibfnamefont{N.}~\bibnamefont{Tamanini}},
  \bibinfo{journal}{Phys. Rev. D} \textbf{\bibinfo{volume}{105}},
  \bibinfo{pages}{064061} (\bibinfo{year}{2022}), \eprint{2109.08748}.

\bibitem[{\citenamefont{Toscani et~al.}(2023)\citenamefont{Toscani, Burke, Liu,
  Zamel, Tamanini, and Pozzoli}}]{Toscani:2023gdf}
\bibinfo{author}{\bibfnamefont{M.}~\bibnamefont{Toscani}},
  \bibinfo{author}{\bibfnamefont{O.}~\bibnamefont{Burke}},
  \bibinfo{author}{\bibfnamefont{C.}~\bibnamefont{Liu}},
  \bibinfo{author}{\bibfnamefont{N.~B.} \bibnamefont{Zamel}},
  \bibinfo{author}{\bibfnamefont{N.}~\bibnamefont{Tamanini}}, \bibnamefont{and}
  \bibinfo{author}{\bibfnamefont{F.}~\bibnamefont{Pozzoli}}
  (\bibinfo{year}{2023}), \eprint{2307.06722}.

\bibitem[{\citenamefont{Speri et~al.}(2021)\citenamefont{Speri, Tamanini,
  Caldwell, Gair, and Wang}}]{Speri:2020hwc}
\bibinfo{author}{\bibfnamefont{L.}~\bibnamefont{Speri}},
  \bibinfo{author}{\bibfnamefont{N.}~\bibnamefont{Tamanini}},
  \bibinfo{author}{\bibfnamefont{R.~R.} \bibnamefont{Caldwell}},
  \bibinfo{author}{\bibfnamefont{J.~R.} \bibnamefont{Gair}}, \bibnamefont{and}
  \bibinfo{author}{\bibfnamefont{B.}~\bibnamefont{Wang}},
  \bibinfo{journal}{Phys. Rev. D} \textbf{\bibinfo{volume}{103}},
  \bibinfo{pages}{083526} (\bibinfo{year}{2021}), \eprint{2010.09049}.

\bibitem[{\citenamefont{Belgacem
  et~al.}(2019{\natexlab{a}})}]{LISACosmologyWorkingGroup:2019mwx}
\bibinfo{author}{\bibfnamefont{E.}~\bibnamefont{Belgacem}} \bibnamefont{et~al.}
  (\bibinfo{collaboration}{LISA Cosmology Working Group}),
  \bibinfo{journal}{JCAP} \textbf{\bibinfo{volume}{07}}, \bibinfo{pages}{024}
  (\bibinfo{year}{2019}{\natexlab{a}}), \eprint{1906.01593}.

\bibitem[{\citenamefont{Cai et~al.}(2017)\citenamefont{Cai, Tamanini, and
  Yang}}]{Cai:2017yww}
\bibinfo{author}{\bibfnamefont{R.-G.} \bibnamefont{Cai}},
  \bibinfo{author}{\bibfnamefont{N.}~\bibnamefont{Tamanini}}, \bibnamefont{and}
  \bibinfo{author}{\bibfnamefont{T.}~\bibnamefont{Yang}},
  \bibinfo{journal}{JCAP} \textbf{\bibinfo{volume}{05}}, \bibinfo{pages}{031}
  (\bibinfo{year}{2017}), \eprint{1703.07323}.

\bibitem[{\citenamefont{Caprini and Tamanini}(2016)}]{Caprini:2016qxs}
\bibinfo{author}{\bibfnamefont{C.}~\bibnamefont{Caprini}} \bibnamefont{and}
  \bibinfo{author}{\bibfnamefont{N.}~\bibnamefont{Tamanini}},
  \bibinfo{journal}{JCAP} \textbf{\bibinfo{volume}{10}}, \bibinfo{pages}{006}
  (\bibinfo{year}{2016}), \eprint{1607.08755}.

\bibitem[{\citenamefont{Auclair
  et~al.}(2022)}]{LISACosmologyWorkingGroup:2022jok}
\bibinfo{author}{\bibfnamefont{P.}~\bibnamefont{Auclair}} \bibnamefont{et~al.}
  (\bibinfo{collaboration}{LISA Cosmology Working Group})
  (\bibinfo{year}{2022}), \eprint{2204.05434}.

\bibitem[{\citenamefont{Saltas et~al.}(2014)\citenamefont{Saltas, Sawicki,
  Amendola, and Kunz}}]{saltas2014anisotropic}
\bibinfo{author}{\bibfnamefont{I.~D.} \bibnamefont{Saltas}},
  \bibinfo{author}{\bibfnamefont{I.}~\bibnamefont{Sawicki}},
  \bibinfo{author}{\bibfnamefont{L.}~\bibnamefont{Amendola}}, \bibnamefont{and}
  \bibinfo{author}{\bibfnamefont{M.}~\bibnamefont{Kunz}},
  \bibinfo{journal}{Physical Review Letters} \textbf{\bibinfo{volume}{113}},
  \bibinfo{pages}{191101} (\bibinfo{year}{2014}).

\bibitem[{\citenamefont{Lombriser and Taylor}(2016)}]{lombriser2016breaking}
\bibinfo{author}{\bibfnamefont{L.}~\bibnamefont{Lombriser}} \bibnamefont{and}
  \bibinfo{author}{\bibfnamefont{A.}~\bibnamefont{Taylor}},
  \bibinfo{journal}{Journal of Cosmology and Astroparticle Physics}
  \textbf{\bibinfo{volume}{2016}}, \bibinfo{pages}{031} (\bibinfo{year}{2016}).

\bibitem[{\citenamefont{Nishizawa}(2018{\natexlab{a}})}]{nishizawa2018generalized}
\bibinfo{author}{\bibfnamefont{A.}~\bibnamefont{Nishizawa}},
  \bibinfo{journal}{Physical Review D} \textbf{\bibinfo{volume}{97}},
  \bibinfo{pages}{104037} (\bibinfo{year}{2018}{\natexlab{a}}).

\bibitem[{\citenamefont{Belgacem
  et~al.}(2018{\natexlab{a}})\citenamefont{Belgacem, Dirian, Foffa, and
  Maggiore}}]{Belgacem:2017ihm}
\bibinfo{author}{\bibfnamefont{E.}~\bibnamefont{Belgacem}},
  \bibinfo{author}{\bibfnamefont{Y.}~\bibnamefont{Dirian}},
  \bibinfo{author}{\bibfnamefont{S.}~\bibnamefont{Foffa}}, \bibnamefont{and}
  \bibinfo{author}{\bibfnamefont{M.}~\bibnamefont{Maggiore}},
  \bibinfo{journal}{Phys. Rev. D} \textbf{\bibinfo{volume}{97}},
  \bibinfo{pages}{104066} (\bibinfo{year}{2018}{\natexlab{a}}),
  \eprint{1712.08108}.

\bibitem[{\citenamefont{Arai and
  Nishizawa}(2018{\natexlab{a}})}]{arai2018generalized}
\bibinfo{author}{\bibfnamefont{S.}~\bibnamefont{Arai}} \bibnamefont{and}
  \bibinfo{author}{\bibfnamefont{A.}~\bibnamefont{Nishizawa}},
  \bibinfo{journal}{Physical Review D} \textbf{\bibinfo{volume}{97}},
  \bibinfo{pages}{104038} (\bibinfo{year}{2018}{\natexlab{a}}).

\bibitem[{\citenamefont{Wolf and Lagos}(2020)}]{wolf2020standard}
\bibinfo{author}{\bibfnamefont{W.~J.} \bibnamefont{Wolf}} \bibnamefont{and}
  \bibinfo{author}{\bibfnamefont{M.}~\bibnamefont{Lagos}},
  \bibinfo{journal}{Physical Review Letters} \textbf{\bibinfo{volume}{124}},
  \bibinfo{pages}{061101} (\bibinfo{year}{2020}).

\bibitem[{\citenamefont{Nishizawa}(2018{\natexlab{b}})}]{Nishizawa:2017nef}
\bibinfo{author}{\bibfnamefont{A.}~\bibnamefont{Nishizawa}},
  \bibinfo{journal}{Phys. Rev. D} \textbf{\bibinfo{volume}{97}},
  \bibinfo{pages}{104037} (\bibinfo{year}{2018}{\natexlab{b}}),
  \eprint{1710.04825}.

\bibitem[{\citenamefont{Arai and Nishizawa}(2018{\natexlab{b}})}]{Arai:2017hxj}
\bibinfo{author}{\bibfnamefont{S.}~\bibnamefont{Arai}} \bibnamefont{and}
  \bibinfo{author}{\bibfnamefont{A.}~\bibnamefont{Nishizawa}},
  \bibinfo{journal}{Phys. Rev. D} \textbf{\bibinfo{volume}{97}},
  \bibinfo{pages}{104038} (\bibinfo{year}{2018}{\natexlab{b}}),
  \eprint{1711.03776}.

\bibitem[{\citenamefont{Belgacem
  et~al.}(2018{\natexlab{b}})\citenamefont{Belgacem, Dirian, Foffa, and
  Maggiore}}]{Belgacem:2018lbp}
\bibinfo{author}{\bibfnamefont{E.}~\bibnamefont{Belgacem}},
  \bibinfo{author}{\bibfnamefont{Y.}~\bibnamefont{Dirian}},
  \bibinfo{author}{\bibfnamefont{S.}~\bibnamefont{Foffa}}, \bibnamefont{and}
  \bibinfo{author}{\bibfnamefont{M.}~\bibnamefont{Maggiore}},
  \bibinfo{journal}{Phys. Rev. D} \textbf{\bibinfo{volume}{98}},
  \bibinfo{pages}{023510} (\bibinfo{year}{2018}{\natexlab{b}}),
  \eprint{1805.08731}.

\bibitem[{\citenamefont{Lagos et~al.}(2019)\citenamefont{Lagos, Fishbach,
  Landry, and Holz}}]{Lagos:2019kds}
\bibinfo{author}{\bibfnamefont{M.}~\bibnamefont{Lagos}},
  \bibinfo{author}{\bibfnamefont{M.}~\bibnamefont{Fishbach}},
  \bibinfo{author}{\bibfnamefont{P.}~\bibnamefont{Landry}}, \bibnamefont{and}
  \bibinfo{author}{\bibfnamefont{D.~E.} \bibnamefont{Holz}},
  \bibinfo{journal}{Phys. Rev. D} \textbf{\bibinfo{volume}{99}},
  \bibinfo{pages}{083504} (\bibinfo{year}{2019}), \eprint{1901.03321}.

\bibitem[{\citenamefont{Zhu et~al.}(2023)\citenamefont{Zhu, Zhao, Yan, Gong,
  and Wang}}]{Zhu:2023wci}
\bibinfo{author}{\bibfnamefont{T.}~\bibnamefont{Zhu}},
  \bibinfo{author}{\bibfnamefont{W.}~\bibnamefont{Zhao}},
  \bibinfo{author}{\bibfnamefont{J.-M.} \bibnamefont{Yan}},
  \bibinfo{author}{\bibfnamefont{C.}~\bibnamefont{Gong}}, \bibnamefont{and}
  \bibinfo{author}{\bibfnamefont{A.}~\bibnamefont{Wang}}
  (\bibinfo{year}{2023}), \eprint{2304.09025}.

\bibitem[{\citenamefont{Abbott
  et~al.}(2021{\natexlab{a}})}]{LIGOScientific:2021djp}
\bibinfo{author}{\bibfnamefont{R.}~\bibnamefont{Abbott}} \bibnamefont{et~al.}
  (\bibinfo{collaboration}{LIGO Scientific, VIRGO, KAGRA})
  (\bibinfo{year}{2021}{\natexlab{a}}), \eprint{2111.03606}.

\bibitem[{\citenamefont{Finke et~al.}(2021{\natexlab{a}})\citenamefont{Finke,
  Foffa, Iacovelli, Maggiore, and Mancarella}}]{Finke:2021aom}
\bibinfo{author}{\bibfnamefont{A.}~\bibnamefont{Finke}},
  \bibinfo{author}{\bibfnamefont{S.}~\bibnamefont{Foffa}},
  \bibinfo{author}{\bibfnamefont{F.}~\bibnamefont{Iacovelli}},
  \bibinfo{author}{\bibfnamefont{M.}~\bibnamefont{Maggiore}}, \bibnamefont{and}
  \bibinfo{author}{\bibfnamefont{M.}~\bibnamefont{Mancarella}},
  \bibinfo{journal}{JCAP} \textbf{\bibinfo{volume}{08}}, \bibinfo{pages}{026}
  (\bibinfo{year}{2021}{\natexlab{a}}), \eprint{2101.12660}.

\bibitem[{\citenamefont{Mancarella
  et~al.}(2022{\natexlab{a}})\citenamefont{Mancarella, Finke, Foffa,
  Genoud-Prachex, Iacovelli, and Maggiore}}]{Mancarella:2022cgn}
\bibinfo{author}{\bibfnamefont{M.}~\bibnamefont{Mancarella}},
  \bibinfo{author}{\bibfnamefont{A.}~\bibnamefont{Finke}},
  \bibinfo{author}{\bibfnamefont{S.}~\bibnamefont{Foffa}},
  \bibinfo{author}{\bibfnamefont{E.}~\bibnamefont{Genoud-Prachex}},
  \bibinfo{author}{\bibfnamefont{F.}~\bibnamefont{Iacovelli}},
  \bibnamefont{and} \bibinfo{author}{\bibfnamefont{M.}~\bibnamefont{Maggiore}},
  in \emph{\bibinfo{booktitle}{{56th Rencontres de Moriond on Gravitation}}}
  (\bibinfo{year}{2022}{\natexlab{a}}), \eprint{2203.09238}.

\bibitem[{\citenamefont{Mastrogiovanni
  et~al.}(2021{\natexlab{b}})\citenamefont{Mastrogiovanni, Haegel,
  Karathanasis, Hernandez, and Steer}}]{Mastrogiovanni:2020mvm}
\bibinfo{author}{\bibfnamefont{S.}~\bibnamefont{Mastrogiovanni}},
  \bibinfo{author}{\bibfnamefont{L.}~\bibnamefont{Haegel}},
  \bibinfo{author}{\bibfnamefont{C.}~\bibnamefont{Karathanasis}},
  \bibinfo{author}{\bibfnamefont{I.~M.~n.} \bibnamefont{Hernandez}},
  \bibnamefont{and} \bibinfo{author}{\bibfnamefont{D.~A.} \bibnamefont{Steer}},
  \bibinfo{journal}{JCAP} \textbf{\bibinfo{volume}{02}}, \bibinfo{pages}{043}
  (\bibinfo{year}{2021}{\natexlab{b}}), \eprint{2010.04047}.

\bibitem[{\citenamefont{de~Rham and Melville}(2018)}]{deRham:2018red}
\bibinfo{author}{\bibfnamefont{C.}~\bibnamefont{de~Rham}} \bibnamefont{and}
  \bibinfo{author}{\bibfnamefont{S.}~\bibnamefont{Melville}},
  \bibinfo{journal}{Phys. Rev. Lett.} \textbf{\bibinfo{volume}{121}},
  \bibinfo{pages}{221101} (\bibinfo{year}{2018}), \eprint{1806.09417}.

\bibitem[{\citenamefont{Abbott
  et~al.}(2017{\natexlab{b}})}]{LIGOScientific:2017zic}
\bibinfo{author}{\bibfnamefont{B.~P.} \bibnamefont{Abbott}}
  \bibnamefont{et~al.} (\bibinfo{collaboration}{LIGO Scientific, Virgo,
  Fermi-GBM, INTEGRAL}), \bibinfo{journal}{Astrophys. J. Lett.}
  \textbf{\bibinfo{volume}{848}}, \bibinfo{pages}{L13}
  (\bibinfo{year}{2017}{\natexlab{b}}), \eprint{1710.05834}.

\bibitem[{\citenamefont{Baker et~al.}(2022)\citenamefont{Baker, Calcagni, Chen,
  Fasiello, Lombriser, Martinovic, Pieroni, Sakellariadou, Tasinato, Bertacca
  et~al.}}]{baker2022measuring}
\bibinfo{author}{\bibfnamefont{T.}~\bibnamefont{Baker}},
  \bibinfo{author}{\bibfnamefont{G.}~\bibnamefont{Calcagni}},
  \bibinfo{author}{\bibfnamefont{A.}~\bibnamefont{Chen}},
  \bibinfo{author}{\bibfnamefont{M.}~\bibnamefont{Fasiello}},
  \bibinfo{author}{\bibfnamefont{L.}~\bibnamefont{Lombriser}},
  \bibinfo{author}{\bibfnamefont{K.}~\bibnamefont{Martinovic}},
  \bibinfo{author}{\bibfnamefont{M.}~\bibnamefont{Pieroni}},
  \bibinfo{author}{\bibfnamefont{M.}~\bibnamefont{Sakellariadou}},
  \bibinfo{author}{\bibfnamefont{G.}~\bibnamefont{Tasinato}},
  \bibinfo{author}{\bibfnamefont{D.}~\bibnamefont{Bertacca}},
  \bibnamefont{et~al.}, \bibinfo{journal}{Journal of Cosmology and
  Astroparticle Physics} \textbf{\bibinfo{volume}{2022}}, \bibinfo{pages}{031}
  (\bibinfo{year}{2022}).

\bibitem[{\citenamefont{Izquierdo-Villalba
  et~al.}(2019)\citenamefont{Izquierdo-Villalba, Angulo, Orsi, Hurier,
  Vilella-Rojo, Bonoli, L{\'o}pez-Sanjuan, Alcaniz, Cenarro,
  Crist{\'o}bal-Hornillos et~al.}}]{villalba19}
\bibinfo{author}{\bibfnamefont{D.}~\bibnamefont{Izquierdo-Villalba}},
  \bibinfo{author}{\bibfnamefont{R.~E.} \bibnamefont{Angulo}},
  \bibinfo{author}{\bibfnamefont{A.}~\bibnamefont{Orsi}},
  \bibinfo{author}{\bibfnamefont{G.}~\bibnamefont{Hurier}},
  \bibinfo{author}{\bibfnamefont{G.}~\bibnamefont{Vilella-Rojo}},
  \bibinfo{author}{\bibfnamefont{S.}~\bibnamefont{Bonoli}},
  \bibinfo{author}{\bibfnamefont{C.}~\bibnamefont{L{\'o}pez-Sanjuan}},
  \bibinfo{author}{\bibfnamefont{J.}~\bibnamefont{Alcaniz}},
  \bibinfo{author}{\bibfnamefont{J.}~\bibnamefont{Cenarro}},
  \bibinfo{author}{\bibfnamefont{D.}~\bibnamefont{Crist{\'o}bal-Hornillos}},
  \bibnamefont{et~al.}, \bibinfo{journal}{Astronomy \& Astrophysics}
  \textbf{\bibinfo{volume}{631}}, \bibinfo{pages}{A82} (\bibinfo{year}{2019}).

\bibitem[{\citenamefont{Barack and Cutler}(2004)}]{barack2004lisa}
\bibinfo{author}{\bibfnamefont{L.}~\bibnamefont{Barack}} \bibnamefont{and}
  \bibinfo{author}{\bibfnamefont{C.}~\bibnamefont{Cutler}},
  \bibinfo{journal}{Physical Review D} \textbf{\bibinfo{volume}{69}},
  \bibinfo{pages}{082005} (\bibinfo{year}{2004}).

\bibitem[{\citenamefont{Henriques et~al.}(2015)\citenamefont{Henriques, White,
  Thomas, Angulo, Guo, Lemson, Springel, and Overzier}}]{henriques2015galaxy}
\bibinfo{author}{\bibfnamefont{B.~M.} \bibnamefont{Henriques}},
  \bibinfo{author}{\bibfnamefont{S.~D.} \bibnamefont{White}},
  \bibinfo{author}{\bibfnamefont{P.~A.} \bibnamefont{Thomas}},
  \bibinfo{author}{\bibfnamefont{R.}~\bibnamefont{Angulo}},
  \bibinfo{author}{\bibfnamefont{Q.}~\bibnamefont{Guo}},
  \bibinfo{author}{\bibfnamefont{G.}~\bibnamefont{Lemson}},
  \bibinfo{author}{\bibfnamefont{V.}~\bibnamefont{Springel}}, \bibnamefont{and}
  \bibinfo{author}{\bibfnamefont{R.}~\bibnamefont{Overzier}},
  \bibinfo{journal}{Monthly Notices of the Royal Astronomical Society}
  \textbf{\bibinfo{volume}{451}}, \bibinfo{pages}{2663} (\bibinfo{year}{2015}).

\bibitem[{\citenamefont{Springel}(2005)}]{springel2005cosmological}
\bibinfo{author}{\bibfnamefont{V.}~\bibnamefont{Springel}},
  \bibinfo{journal}{Monthly notices of the royal astronomical society}
  \textbf{\bibinfo{volume}{364}}, \bibinfo{pages}{1105} (\bibinfo{year}{2005}).

\bibitem[{\citenamefont{Muttoni et~al.}(2023)\citenamefont{Muttoni, Laghi,
  Tamanini, Marsat, and Izquierdo-Villalba}}]{muttoni2023dark}
\bibinfo{author}{\bibfnamefont{N.}~\bibnamefont{Muttoni}},
  \bibinfo{author}{\bibfnamefont{D.}~\bibnamefont{Laghi}},
  \bibinfo{author}{\bibfnamefont{N.}~\bibnamefont{Tamanini}},
  \bibinfo{author}{\bibfnamefont{S.}~\bibnamefont{Marsat}}, \bibnamefont{and}
  \bibinfo{author}{\bibfnamefont{D.}~\bibnamefont{Izquierdo-Villalba}},
  \bibinfo{journal}{arXiv preprint arXiv:2303.10693}  (\bibinfo{year}{2023}).

\bibitem[{\citenamefont{Muttoni et~al.}(2022)\citenamefont{Muttoni, Mangiagli,
  Sesana, Laghi, Del~Pozzo, Izquierdo-Villalba, and Rosati}}]{Muttoni:2021veo}
\bibinfo{author}{\bibfnamefont{N.}~\bibnamefont{Muttoni}},
  \bibinfo{author}{\bibfnamefont{A.}~\bibnamefont{Mangiagli}},
  \bibinfo{author}{\bibfnamefont{A.}~\bibnamefont{Sesana}},
  \bibinfo{author}{\bibfnamefont{D.}~\bibnamefont{Laghi}},
  \bibinfo{author}{\bibfnamefont{W.}~\bibnamefont{Del~Pozzo}},
  \bibinfo{author}{\bibfnamefont{D.}~\bibnamefont{Izquierdo-Villalba}},
  \bibnamefont{and} \bibinfo{author}{\bibfnamefont{M.}~\bibnamefont{Rosati}},
  \bibinfo{journal}{Phys. Rev. D} \textbf{\bibinfo{volume}{105}},
  \bibinfo{pages}{043509} (\bibinfo{year}{2022}), \eprint{2109.13934}.

\bibitem[{\citenamefont{Del~Pozzo}(2012)}]{DelPozzo:2011vcw}
\bibinfo{author}{\bibfnamefont{W.}~\bibnamefont{Del~Pozzo}},
  \bibinfo{journal}{Phys. Rev. D} \textbf{\bibinfo{volume}{86}},
  \bibinfo{pages}{043011} (\bibinfo{year}{2012}), \eprint{1108.1317}.

\bibitem[{\citenamefont{{Del Pozzo} and Laghi}(2022)}]{cosmoLISA}
\bibinfo{author}{\bibfnamefont{W.}~\bibnamefont{{Del Pozzo}}} \bibnamefont{and}
  \bibinfo{author}{\bibfnamefont{D.}~\bibnamefont{Laghi}},
  \emph{\bibinfo{title}{wdpozzo/cosmolisa}} (\bibinfo{year}{2022}),
  \urlprefix\url{https://github.com/wdpozzo/cosmolisa}.

\bibitem[{\citenamefont{{Del Pozzo} and Veitch}(2021)}]{raynest}
\bibinfo{author}{\bibfnamefont{W.}~\bibnamefont{{Del Pozzo}}} \bibnamefont{and}
  \bibinfo{author}{\bibfnamefont{J.}~\bibnamefont{Veitch}},
  \emph{\bibinfo{title}{wdpozzo/raynest}} (\bibinfo{year}{2021}),
  \urlprefix\url{https://github.com/wdpozzo/raynest}.

\bibitem[{ray()}]{ray}
\urlprefix\url{https://pypi.org/project/ray/}.

\bibitem[{\citenamefont{Mandel et~al.}(2019)\citenamefont{Mandel, Farr, and
  Gair}}]{mandel2019extracting}
\bibinfo{author}{\bibfnamefont{I.}~\bibnamefont{Mandel}},
  \bibinfo{author}{\bibfnamefont{W.~M.} \bibnamefont{Farr}}, \bibnamefont{and}
  \bibinfo{author}{\bibfnamefont{J.~R.} \bibnamefont{Gair}},
  \bibinfo{journal}{Monthly Notices of the Royal Astronomical Society}
  \textbf{\bibinfo{volume}{486}}, \bibinfo{pages}{1086} (\bibinfo{year}{2019}).

\bibitem[{\citenamefont{Vitale et~al.}(2022)\citenamefont{Vitale, Gerosa, Farr,
  and Taylor}}]{vitale2022inferring}
\bibinfo{author}{\bibfnamefont{S.}~\bibnamefont{Vitale}},
  \bibinfo{author}{\bibfnamefont{D.}~\bibnamefont{Gerosa}},
  \bibinfo{author}{\bibfnamefont{W.~M.} \bibnamefont{Farr}}, \bibnamefont{and}
  \bibinfo{author}{\bibfnamefont{S.~R.} \bibnamefont{Taylor}}, in
  \emph{\bibinfo{booktitle}{Handbook of Gravitational Wave Astronomy}}
  (\bibinfo{publisher}{Springer}, \bibinfo{year}{2022}), pp.
  \bibinfo{pages}{1--60}.

\bibitem[{\citenamefont{Maggiore}(2017)}]{maggiore2017nonlocal}
\bibinfo{author}{\bibfnamefont{M.}~\bibnamefont{Maggiore}}, in
  \emph{\bibinfo{booktitle}{Gravity and the Quantum: Pedagogical Essays on
  Cosmology, Astrophysics, and Quantum Gravity}}
  (\bibinfo{publisher}{Springer}, \bibinfo{year}{2017}), pp.
  \bibinfo{pages}{221--281}.

\bibitem[{\citenamefont{Belgacem
  et~al.}(2019{\natexlab{b}})\citenamefont{Belgacem, Dirian, Foffa, Howell,
  Maggiore, and Regimbau}}]{Belgacem:2019tbw}
\bibinfo{author}{\bibfnamefont{E.}~\bibnamefont{Belgacem}},
  \bibinfo{author}{\bibfnamefont{Y.}~\bibnamefont{Dirian}},
  \bibinfo{author}{\bibfnamefont{S.}~\bibnamefont{Foffa}},
  \bibinfo{author}{\bibfnamefont{E.~J.} \bibnamefont{Howell}},
  \bibinfo{author}{\bibfnamefont{M.}~\bibnamefont{Maggiore}}, \bibnamefont{and}
  \bibinfo{author}{\bibfnamefont{T.}~\bibnamefont{Regimbau}},
  \bibinfo{journal}{JCAP} \textbf{\bibinfo{volume}{08}}, \bibinfo{pages}{015}
  (\bibinfo{year}{2019}{\natexlab{b}}), \eprint{1907.01487}.

\bibitem[{\citenamefont{{Virgo Millennium Database}}(2012)}]{VMD}
\bibinfo{author}{\bibnamefont{{Virgo Millennium Database}}}
  (\bibinfo{year}{2012}),
  \urlprefix\url{http://gavo.mpa-garching.mpg.de/Millennium/Help?page=databases/henriques2012a/database}.

\bibitem[{\citenamefont{Abbott et~al.}(2019)\citenamefont{Abbott, Abbott,
  Abbott, Acernese, Ackley, Adams, Adams, Addesso, Adhikari, Adya
  et~al.}}]{abbott2019tests}
\bibinfo{author}{\bibfnamefont{B.~P.} \bibnamefont{Abbott}},
  \bibinfo{author}{\bibfnamefont{R.}~\bibnamefont{Abbott}},
  \bibinfo{author}{\bibfnamefont{T.}~\bibnamefont{Abbott}},
  \bibinfo{author}{\bibfnamefont{F.}~\bibnamefont{Acernese}},
  \bibinfo{author}{\bibfnamefont{K.}~\bibnamefont{Ackley}},
  \bibinfo{author}{\bibfnamefont{C.}~\bibnamefont{Adams}},
  \bibinfo{author}{\bibfnamefont{T.}~\bibnamefont{Adams}},
  \bibinfo{author}{\bibfnamefont{P.}~\bibnamefont{Addesso}},
  \bibinfo{author}{\bibfnamefont{R.~X.} \bibnamefont{Adhikari}},
  \bibinfo{author}{\bibfnamefont{V.~B.} \bibnamefont{Adya}},
  \bibnamefont{et~al.}, \bibinfo{journal}{Physical review letters}
  \textbf{\bibinfo{volume}{123}}, \bibinfo{pages}{011102}
  (\bibinfo{year}{2019}).

\bibitem[{\citenamefont{Abbott et~al.}(2021{\natexlab{b}})\citenamefont{Abbott,
  Abe, Acernese, Ackley, Adhikari, Adhikari, Adkins, Adya, Affeldt, Agarwal
  et~al.}}]{abbott2021tests}
\bibinfo{author}{\bibfnamefont{R.}~\bibnamefont{Abbott}},
  \bibinfo{author}{\bibfnamefont{H.}~\bibnamefont{Abe}},
  \bibinfo{author}{\bibfnamefont{F.}~\bibnamefont{Acernese}},
  \bibinfo{author}{\bibfnamefont{K.}~\bibnamefont{Ackley}},
  \bibinfo{author}{\bibfnamefont{N.}~\bibnamefont{Adhikari}},
  \bibinfo{author}{\bibfnamefont{R.}~\bibnamefont{Adhikari}},
  \bibinfo{author}{\bibfnamefont{V.}~\bibnamefont{Adkins}},
  \bibinfo{author}{\bibfnamefont{V.}~\bibnamefont{Adya}},
  \bibinfo{author}{\bibfnamefont{C.}~\bibnamefont{Affeldt}},
  \bibinfo{author}{\bibfnamefont{D.}~\bibnamefont{Agarwal}},
  \bibnamefont{et~al.}, \bibinfo{journal}{arXiv preprint arXiv:2112.06861}
  (\bibinfo{year}{2021}{\natexlab{b}}).

\bibitem[{\citenamefont{D{\'a}lya et~al.}(2018)\citenamefont{D{\'a}lya,
  Galg{\'o}czi, Dobos, Frei, Heng, Macas, Messenger, Raffai, and
  de~Souza}}]{dalya2018glade}
\bibinfo{author}{\bibfnamefont{G.}~\bibnamefont{D{\'a}lya}},
  \bibinfo{author}{\bibfnamefont{G.}~\bibnamefont{Galg{\'o}czi}},
  \bibinfo{author}{\bibfnamefont{L.}~\bibnamefont{Dobos}},
  \bibinfo{author}{\bibfnamefont{Z.}~\bibnamefont{Frei}},
  \bibinfo{author}{\bibfnamefont{I.~S.} \bibnamefont{Heng}},
  \bibinfo{author}{\bibfnamefont{R.}~\bibnamefont{Macas}},
  \bibinfo{author}{\bibfnamefont{C.}~\bibnamefont{Messenger}},
  \bibinfo{author}{\bibfnamefont{P.}~\bibnamefont{Raffai}}, \bibnamefont{and}
  \bibinfo{author}{\bibfnamefont{R.~S.} \bibnamefont{de~Souza}},
  \bibinfo{journal}{Monthly Notices of the Royal Astronomical Society}
  \textbf{\bibinfo{volume}{479}}, \bibinfo{pages}{2374} (\bibinfo{year}{2018}).

\bibitem[{\citenamefont{Mancarella
  et~al.}(2022{\natexlab{b}})\citenamefont{Mancarella, Genoud-Prachex, and
  Maggiore}}]{Mancarella:2021ecn}
\bibinfo{author}{\bibfnamefont{M.}~\bibnamefont{Mancarella}},
  \bibinfo{author}{\bibfnamefont{E.}~\bibnamefont{Genoud-Prachex}},
  \bibnamefont{and} \bibinfo{author}{\bibfnamefont{M.}~\bibnamefont{Maggiore}},
  \bibinfo{journal}{Phys. Rev. D} \textbf{\bibinfo{volume}{105}},
  \bibinfo{pages}{064030} (\bibinfo{year}{2022}{\natexlab{b}}),
  \eprint{2112.05728}.

\bibitem[{\citenamefont{Leyde et~al.}(2022)\citenamefont{Leyde, Mastrogiovanni,
  Steer, Chassande-Mottin, and Karathanasis}}]{Leyde:2022orh}
\bibinfo{author}{\bibfnamefont{K.}~\bibnamefont{Leyde}},
  \bibinfo{author}{\bibfnamefont{S.}~\bibnamefont{Mastrogiovanni}},
  \bibinfo{author}{\bibfnamefont{D.~A.} \bibnamefont{Steer}},
  \bibinfo{author}{\bibfnamefont{E.}~\bibnamefont{Chassande-Mottin}},
  \bibnamefont{and}
  \bibinfo{author}{\bibfnamefont{C.}~\bibnamefont{Karathanasis}},
  \bibinfo{journal}{JCAP} \textbf{\bibinfo{volume}{09}}, \bibinfo{pages}{012}
  (\bibinfo{year}{2022}), \eprint{2202.00025}.

\bibitem[{\citenamefont{Chen et~al.}(2023)\citenamefont{Chen, Gray, and
  Baker}}]{chen2023testing}
\bibinfo{author}{\bibfnamefont{A.}~\bibnamefont{Chen}},
  \bibinfo{author}{\bibfnamefont{R.}~\bibnamefont{Gray}}, \bibnamefont{and}
  \bibinfo{author}{\bibfnamefont{T.}~\bibnamefont{Baker}},
  \bibinfo{journal}{arXiv preprint arXiv:2309.03833}  (\bibinfo{year}{2023}).

\bibitem[{\citenamefont{D{\'a}lya et~al.}(2022)\citenamefont{D{\'a}lya,
  D{\'\i}az, Bouchet, Frei, Jasche, Lavaux, Macas, Mukherjee, P{\'a}lfi,
  De~Souza et~al.}}]{dalya2022glade+}
\bibinfo{author}{\bibfnamefont{G.}~\bibnamefont{D{\'a}lya}},
  \bibinfo{author}{\bibfnamefont{R.}~\bibnamefont{D{\'\i}az}},
  \bibinfo{author}{\bibfnamefont{F.}~\bibnamefont{Bouchet}},
  \bibinfo{author}{\bibfnamefont{Z.}~\bibnamefont{Frei}},
  \bibinfo{author}{\bibfnamefont{J.}~\bibnamefont{Jasche}},
  \bibinfo{author}{\bibfnamefont{G.}~\bibnamefont{Lavaux}},
  \bibinfo{author}{\bibfnamefont{R.}~\bibnamefont{Macas}},
  \bibinfo{author}{\bibfnamefont{S.}~\bibnamefont{Mukherjee}},
  \bibinfo{author}{\bibfnamefont{M.}~\bibnamefont{P{\'a}lfi}},
  \bibinfo{author}{\bibfnamefont{R.}~\bibnamefont{De~Souza}},
  \bibnamefont{et~al.}, \bibinfo{journal}{Monthly Notices of the Royal
  Astronomical Society} \textbf{\bibinfo{volume}{514}}, \bibinfo{pages}{1403}
  (\bibinfo{year}{2022}).

\bibitem[{\citenamefont{Mukherjee et~al.}(2021)\citenamefont{Mukherjee,
  Wandelt, and Silk}}]{mukherjee2021testing}
\bibinfo{author}{\bibfnamefont{S.}~\bibnamefont{Mukherjee}},
  \bibinfo{author}{\bibfnamefont{B.~D.} \bibnamefont{Wandelt}},
  \bibnamefont{and} \bibinfo{author}{\bibfnamefont{J.}~\bibnamefont{Silk}},
  \bibinfo{journal}{Monthly Notices of the Royal Astronomical Society}
  \textbf{\bibinfo{volume}{502}}, \bibinfo{pages}{1136} (\bibinfo{year}{2021}).

\bibitem[{\citenamefont{Branchesi et~al.}(2023)\citenamefont{Branchesi,
  Maggiore, Alonso, Badger, Banerjee, Beirnaert, Belgacem, Bhagwat, Boileau,
  Borhanian et~al.}}]{branchesi2023science}
\bibinfo{author}{\bibfnamefont{M.}~\bibnamefont{Branchesi}},
  \bibinfo{author}{\bibfnamefont{M.}~\bibnamefont{Maggiore}},
  \bibinfo{author}{\bibfnamefont{D.}~\bibnamefont{Alonso}},
  \bibinfo{author}{\bibfnamefont{C.}~\bibnamefont{Badger}},
  \bibinfo{author}{\bibfnamefont{B.}~\bibnamefont{Banerjee}},
  \bibinfo{author}{\bibfnamefont{F.}~\bibnamefont{Beirnaert}},
  \bibinfo{author}{\bibfnamefont{E.}~\bibnamefont{Belgacem}},
  \bibinfo{author}{\bibfnamefont{S.}~\bibnamefont{Bhagwat}},
  \bibinfo{author}{\bibfnamefont{G.}~\bibnamefont{Boileau}},
  \bibinfo{author}{\bibfnamefont{S.}~\bibnamefont{Borhanian}},
  \bibnamefont{et~al.}, \bibinfo{journal}{Journal of Cosmology and
  Astroparticle Physics} \textbf{\bibinfo{volume}{2023}}, \bibinfo{pages}{068}
  (\bibinfo{year}{2023}).

\bibitem[{\citenamefont{Hild et~al.}(2008)\citenamefont{Hild, Chelkowski, and
  Freise}}]{hild2008pushing}
\bibinfo{author}{\bibfnamefont{S.}~\bibnamefont{Hild}},
  \bibinfo{author}{\bibfnamefont{S.}~\bibnamefont{Chelkowski}},
  \bibnamefont{and} \bibinfo{author}{\bibfnamefont{A.}~\bibnamefont{Freise}},
  \bibinfo{journal}{arXiv preprint arXiv:0810.0604}  (\bibinfo{year}{2008}).

\bibitem[{\citenamefont{Punturo et~al.}(2010)\citenamefont{Punturo, Abernathy,
  Acernese, Allen, Andersson, Arun, Barone, Barr, Barsuglia, Beker
  et~al.}}]{punturo2010einstein}
\bibinfo{author}{\bibfnamefont{M.}~\bibnamefont{Punturo}},
  \bibinfo{author}{\bibfnamefont{M.}~\bibnamefont{Abernathy}},
  \bibinfo{author}{\bibfnamefont{F.}~\bibnamefont{Acernese}},
  \bibinfo{author}{\bibfnamefont{B.}~\bibnamefont{Allen}},
  \bibinfo{author}{\bibfnamefont{N.}~\bibnamefont{Andersson}},
  \bibinfo{author}{\bibfnamefont{K.}~\bibnamefont{Arun}},
  \bibinfo{author}{\bibfnamefont{F.}~\bibnamefont{Barone}},
  \bibinfo{author}{\bibfnamefont{B.}~\bibnamefont{Barr}},
  \bibinfo{author}{\bibfnamefont{M.}~\bibnamefont{Barsuglia}},
  \bibinfo{author}{\bibfnamefont{M.}~\bibnamefont{Beker}},
  \bibnamefont{et~al.}, \bibinfo{journal}{Classical and Quantum Gravity}
  \textbf{\bibinfo{volume}{27}}, \bibinfo{pages}{194002}
  (\bibinfo{year}{2010}).

\bibitem[{\citenamefont{Hild et~al.}(2011)\citenamefont{Hild, Abernathy,
  Acernese, Amaro-Seoane, Andersson, Arun, Barone, Barr, Barsuglia, Beker
  et~al.}}]{hild2011sensitivity}
\bibinfo{author}{\bibfnamefont{S.}~\bibnamefont{Hild}},
  \bibinfo{author}{\bibfnamefont{M.}~\bibnamefont{Abernathy}},
  \bibinfo{author}{\bibfnamefont{F.~e.} \bibnamefont{Acernese}},
  \bibinfo{author}{\bibfnamefont{P.}~\bibnamefont{Amaro-Seoane}},
  \bibinfo{author}{\bibfnamefont{N.}~\bibnamefont{Andersson}},
  \bibinfo{author}{\bibfnamefont{K.}~\bibnamefont{Arun}},
  \bibinfo{author}{\bibfnamefont{F.}~\bibnamefont{Barone}},
  \bibinfo{author}{\bibfnamefont{B.}~\bibnamefont{Barr}},
  \bibinfo{author}{\bibfnamefont{M.}~\bibnamefont{Barsuglia}},
  \bibinfo{author}{\bibfnamefont{M.}~\bibnamefont{Beker}},
  \bibnamefont{et~al.}, \bibinfo{journal}{Classical and Quantum gravity}
  \textbf{\bibinfo{volume}{28}}, \bibinfo{pages}{094013}
  (\bibinfo{year}{2011}).

\bibitem[{\citenamefont{Jiang and Yagi}(2021)}]{Jiang:2021mpd}
\bibinfo{author}{\bibfnamefont{N.}~\bibnamefont{Jiang}} \bibnamefont{and}
  \bibinfo{author}{\bibfnamefont{K.}~\bibnamefont{Yagi}},
  \bibinfo{journal}{Phys. Rev. D} \textbf{\bibinfo{volume}{103}},
  \bibinfo{pages}{124047} (\bibinfo{year}{2021}), \eprint{2104.04442}.

\bibitem[{\citenamefont{Kawamura et~al.}(2011)\citenamefont{Kawamura, Ando,
  Seto, Sato, Nakamura, Tsubono, Kanda, Tanaka, Yokoyama, Funaki
  et~al.}}]{kawamura2011japanese}
\bibinfo{author}{\bibfnamefont{S.}~\bibnamefont{Kawamura}},
  \bibinfo{author}{\bibfnamefont{M.}~\bibnamefont{Ando}},
  \bibinfo{author}{\bibfnamefont{N.}~\bibnamefont{Seto}},
  \bibinfo{author}{\bibfnamefont{S.}~\bibnamefont{Sato}},
  \bibinfo{author}{\bibfnamefont{T.}~\bibnamefont{Nakamura}},
  \bibinfo{author}{\bibfnamefont{K.}~\bibnamefont{Tsubono}},
  \bibinfo{author}{\bibfnamefont{N.}~\bibnamefont{Kanda}},
  \bibinfo{author}{\bibfnamefont{T.}~\bibnamefont{Tanaka}},
  \bibinfo{author}{\bibfnamefont{J.}~\bibnamefont{Yokoyama}},
  \bibinfo{author}{\bibfnamefont{I.}~\bibnamefont{Funaki}},
  \bibnamefont{et~al.}, \bibinfo{journal}{Classical and Quantum Gravity}
  \textbf{\bibinfo{volume}{28}}, \bibinfo{pages}{094011}
  (\bibinfo{year}{2011}).

\bibitem[{\citenamefont{Kawamura et~al.}(2021)\citenamefont{Kawamura, Ando,
  Seto, Sato, Musha, Kawano, Yokoyama, Tanaka, Ioka, Akutsu
  et~al.}}]{kawamura2021current}
\bibinfo{author}{\bibfnamefont{S.}~\bibnamefont{Kawamura}},
  \bibinfo{author}{\bibfnamefont{M.}~\bibnamefont{Ando}},
  \bibinfo{author}{\bibfnamefont{N.}~\bibnamefont{Seto}},
  \bibinfo{author}{\bibfnamefont{S.}~\bibnamefont{Sato}},
  \bibinfo{author}{\bibfnamefont{M.}~\bibnamefont{Musha}},
  \bibinfo{author}{\bibfnamefont{I.}~\bibnamefont{Kawano}},
  \bibinfo{author}{\bibfnamefont{J.}~\bibnamefont{Yokoyama}},
  \bibinfo{author}{\bibfnamefont{T.}~\bibnamefont{Tanaka}},
  \bibinfo{author}{\bibfnamefont{K.}~\bibnamefont{Ioka}},
  \bibinfo{author}{\bibfnamefont{T.}~\bibnamefont{Akutsu}},
  \bibnamefont{et~al.}, \bibinfo{journal}{Progress of Theoretical and
  Experimental Physics} \textbf{\bibinfo{volume}{2021}},
  \bibinfo{pages}{05A105} (\bibinfo{year}{2021}).

\bibitem[{\citenamefont{Finke et~al.}(2021{\natexlab{b}})\citenamefont{Finke,
  Foffa, Iacovelli, Maggiore, and Mancarella}}]{Finke:2021znb}
\bibinfo{author}{\bibfnamefont{A.}~\bibnamefont{Finke}},
  \bibinfo{author}{\bibfnamefont{S.}~\bibnamefont{Foffa}},
  \bibinfo{author}{\bibfnamefont{F.}~\bibnamefont{Iacovelli}},
  \bibinfo{author}{\bibfnamefont{M.}~\bibnamefont{Maggiore}}, \bibnamefont{and}
  \bibinfo{author}{\bibfnamefont{M.}~\bibnamefont{Mancarella}},
  \bibinfo{journal}{Phys. Rev. D} \textbf{\bibinfo{volume}{104}},
  \bibinfo{pages}{084057} (\bibinfo{year}{2021}{\natexlab{b}}),
  \eprint{2107.05046}.

\bibitem[{\citenamefont{Matos et~al.}(2021)\citenamefont{Matos, Calv{\~a}o, and
  Waga}}]{matos2021gravitational}
\bibinfo{author}{\bibfnamefont{I.~S.} \bibnamefont{Matos}},
  \bibinfo{author}{\bibfnamefont{M.~O.} \bibnamefont{Calv{\~a}o}},
  \bibnamefont{and} \bibinfo{author}{\bibfnamefont{I.}~\bibnamefont{Waga}},
  \bibinfo{journal}{Physical Review D} \textbf{\bibinfo{volume}{103}},
  \bibinfo{pages}{104059} (\bibinfo{year}{2021}).

\bibitem[{\citenamefont{Yang}(2021)}]{Yang:2021qge}
\bibinfo{author}{\bibfnamefont{T.}~\bibnamefont{Yang}}, \bibinfo{journal}{JCAP}
  \textbf{\bibinfo{volume}{05}}, \bibinfo{pages}{044} (\bibinfo{year}{2021}),
  \eprint{2103.01923}.

\bibitem[{\citenamefont{Tamanini}(2017)}]{Tamanini:2016uin}
\bibinfo{author}{\bibfnamefont{N.}~\bibnamefont{Tamanini}},
  \bibinfo{journal}{J. Phys. Conf. Ser.} \textbf{\bibinfo{volume}{840}},
  \bibinfo{pages}{012029} (\bibinfo{year}{2017}), \eprint{1612.02634}.

\end{thebibliography}

\end{document}